\documentclass[11pt, titlepage]{article}
\usepackage[sort&compress, round, semicolon, authoryear]{natbib}
\usepackage{graphicx, lscape}
\usepackage{graphics, color}
\usepackage{epsfig}
\usepackage{epstopdf}
\usepackage{wrapfig}
\usepackage{caption}
\usepackage{subcaption}
\usepackage{amsmath, amsthm, amssymb, amscd}
\usepackage{latexsym}
\usepackage{multirow} 
\usepackage{array}
\usepackage{color}
\usepackage{rotating}
\usepackage[hyphens]{url}
\usepackage[hidelinks]{hyperref}  
\usepackage{cleveref}
\usepackage{fullpage}
\usepackage{fancyvrb}
\usepackage{pdfpages}
\usepackage{fmtcount}
\usepackage[margin=3.3cm]{geometry}
\usepackage{verbatim}
\usepackage[english]{babel}
\usepackage[T1]{fontenc}
\usepackage[utf8]{inputenc}
\usepackage{tikz} 
\usetikzlibrary{arrows, decorations.pathmorphing, backgrounds, fit, positioning, shapes.symbols, chains}
\usetikzlibrary{decorations.markings}
\usepackage{lscape}
\usepackage{algorithm}
\usepackage{algpseudocode}
\usepackage{soul}
\usepackage{booktabs}
\usepackage{enumerate}
\usepackage{colortbl}
\long\def\comment#1{} 
\usepackage{booktabs}
\usepackage{yhmath}  
\usepackage{mathtools}
\usepackage{rotating}
\usepackage{bm,upgreek}
\usepackage{mathrsfs}
\usepackage{changepage}
\usepackage{enumitem}
\usepackage{comment}
\usepackage[title,titletoc]{appendix}
\usepackage{accents}
\usepackage{chngcntr}
\newlength{\dhatheight}

\makeatletter
\newcounter{phase}[algorithm]
\newlength{\phaserulewidth}
\newcommand{\setphaserulewidth}{\setlength{\phaserulewidth}}
\newcommand{\phase}[1]{%
	\vspace{-1.25ex}
	\Statex\leavevmode\llap{\rule{\dimexpr\labelwidth+\labelsep}{\phaserulewidth}}\rule{\linewidth}{\phaserulewidth}
	\Statex\strut\refstepcounter{phase}\textit{Step~\thephase~--~#1}
	\vspace{-1.25ex}\Statex\leavevmode\llap{\rule{\dimexpr\labelwidth+\labelsep}{\phaserulewidth}}\rule{\linewidth}{\phaserulewidth}}
\makeatother
\setphaserulewidth{.7pt}

\usepackage{stackengine}

\stackMath
\newcommand\tenq[2][1]{%
	\def\useanchorwidth{T}%
	\ifnum#1>1%
	\stackunder[0pt]{\tenq[\numexpr#1-1\relax]{#2}}{\scriptscriptstyle\sim}%
	\else%
	\stackunder[1pt]{#2}{\scriptscriptstyle\sim}%
	\fi%
}

\usepackage{xcolor}
\definecolor{jm}{rgb}{0.5, 0.1, 0.3}
\definecolor{jh}{rgb}{0.2, 0.1, 0.6}
\definecolor{dark_green}{rgb}{0.1, 0.5, 0.1}

\newcommand{\romnum}[1]{\lowercase\expandafter{\romannumeral #1\relax}}

\newtheorem{proposition}{Proposition}[section]
\newtheorem{corollary}{Corollary}[section]
\newtheorem{theorem}{Theorem}[section]

\theoremstyle{definition}
\newtheorem{definition}{Definition}[section]

\usepackage{mathtools}

\newcommand{\bdm}{\begin{displaymath}}
	\newcommand{\edm}{\end{displaymath}}

\newcommand{\mc}{\mathcal}
\newcommand{\mr}{\mathrm}

\begin{document}
	
	\newpage
	\begin{center}
		{\Large\bf Graph Canonical Coherence Analysis
			\medskip
		}
		\vskip 7mm
		
		\renewcommand{\thefootnote}{\fnsymbol{footnote}}
		{\large\sc Kyusoon Kim$^{1}$ and Hee-Seok Oh$^{2}$\footnote{Corresponding author: heeseok@stats.snu.ac.kr}}\\
		{\large
			$^{1}$Department of Statistics and Actuarial Science, Soongsil University, Seoul 06978, Korea\\
			$^{2}$Department of Statistics, Seoul National University, Seoul 08826, Korea
		}
	\end{center}
	\vskip 5mm
	
	\begin{quote}
		\begin{center}
			\textbf{Abstract}
		\end{center}
		
		We propose graph canonical coherence analysis (gCChA), a novel framework that extends canonical correlation analysis to multivariate graph signals in the graph frequency domain. The proposed method addresses challenges posed by the inherent features of graphs: discreteness, finiteness, and irregularity. It identifies pairs of canonical graph signals that maximize their coherence, enabling the exploration of relationships between two sets of graph signals from a spectral perspective. This framework shows how these relationships change across different structural scales of the graph. We demonstrate the usefulness of this method through a simulation study and an application to economic and energy datasets of G20 countries.
		
		\end{quote}

	\pagenumbering{arabic}
	
	
	\section{Introduction} \label{sec:intro}
	Canonical correlation analysis (CCA), originally proposed by \cite{hotelling1935most, hotelling1936relation}, is a classical multivariate method for quantifying a linear relationship between two sets of multidimensional variables. This method finds pairs of canonical components, which are linear combinations of variables from each set that achieve the highest possible correlation. Through these components, CCA summarizes the dependence structure between the two sets in a lower-dimensional space. In this regard, it can be viewed as a dimension reduction method that extends the concept of principal component analysis (PCA) to two sets of variables. Thus, CCA is often used to evaluate both the strength and the underlying structure of relationships between the two sets. It has been applied in a wide range of fields, including neuroscience \citep{zhuang2020technical}, bioinformatics \citep{wrobel2024data}, finance \citep{mazuruse2014canonical}, and deep learning \citep{andrew13}, in tasks such as blind source separation, clustering, and classification. For a comprehensive overview of CCA, refer to \cite{yang2019survey}, \cite{johnson2007applied}, and \cite{hardoon2004canonical}.
	
\begin{figure}[!t]
		\centering
		\includegraphics[width=0.61\textwidth]{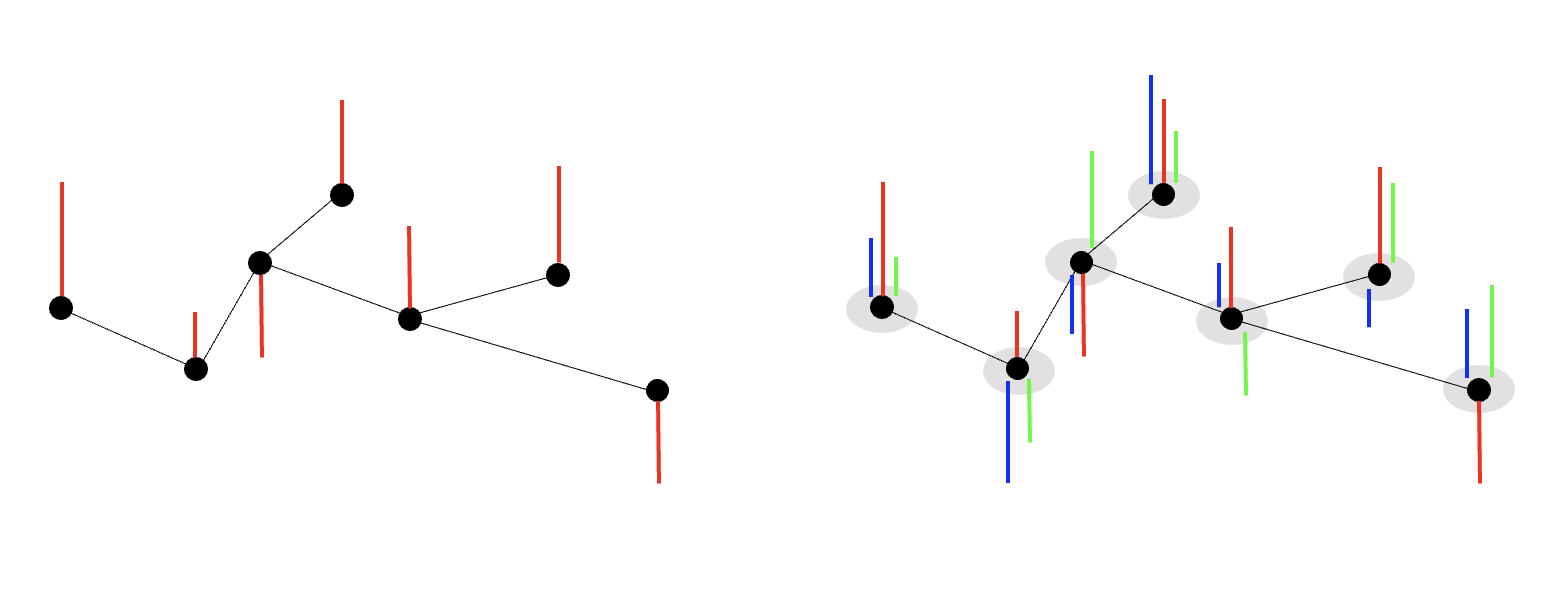}
		\vspace{-1.7em}
		\caption{Illustration of univariate (left) and multivariate (right) graph signals.}
		\label{fig:univ_multiv_graphsignals}
	\end{figure}

	Classical CCA deals with multidimensional data in Euclidean space, where each set of variables is represented as a vector. \cite{Brillinger2001time} further extended CCA to analyze two multivariate time series by formulating it in the frequency domain. However, in many fields, data are observed on irregular domains with graph structures, known as graph signals \citep{leus2023graph, shuman2013emerging}, highlighting the need to develop CCA tailored to such domains. Specifically, univariate graph signals are scalar values located on the nodes of a graph, while multivariate graph signals consist of vector-valued data associated with each node, as shown in \Cref{fig:univ_multiv_graphsignals}. In the figure, the height and direction (upward or downward) of each bar indicate the magnitude and sign of the signal value mapped to each node, and bars with the same color correspond to the same attribute in the multivariate case. Graph signals defined on a path or cycle graph can be viewed as special cases that correspond to time series.
	
	Developing CCA for multivariate graph signals presents unique challenges due to the inherent discreteness, finiteness, and irregularity of graphs. In this study, we propose a new CCA framework for graph signals called \textit{graph canonical coherence analysis (gCChA)}. This framework aims to discover relationships between two multivariate graph signals on a common underlying graph. To account for the distinctive characteristics of the graph domain, we develop a framework in the graph frequency domain that bridges time-domain and graph-domain analyses, drawing on the frequency-domain approach of \citet{Brillinger2001time}. Unlike several previous works that operate in the node domain,  gCChA provides a systematic way to describe the relationships between two multivariate graph signals using low-dimensional canonical graph signals and to analyze these relationships from a spectral perspective across graph frequencies.
	
	The \textsf{R} codes and datasets used in the numerical experiments are available at \url{https://github.com/qsoon/gCChA}.
	
	\section{Background} \label{sec:prelim}
	\subsection{Graph Signals and Graph Frequency Domain} \label{sec:graphsbasic}
	Let $\mc{G} = (\mc{V}, E, W)$ denote a weighted graph, possibly directed, where $\mc{V}$ is the set of $n$ nodes, $E$ is the set of edges, and $W$ is the weighted adjacency matrix. We assume that $\mc{G}$ is finite, connected, and simple, meaning it has no self-loops or multiple edges. The nonnegative element $w_{ij}$ of $W$ represents the weight of the edge between nodes $i$ and $j$, and a value of $w_{ij}=0$ indicates no direct connection between them. A graph signal assigns a scalar or vector to each node of $\mc{G}$. A univariate graph signal can be viewed as a function $x: \mc{V} \rightarrow \mathbb{R}$ (or $\mathbb{C}$), while a multivariate graph signal takes values in higher-dimensional codomains. In the univariate case, the signal is represented by a vector $x$ of size $n$, whose $i$th entry corresponds to the value at node $i$.
	
	A key concept in graph signal processing (GSP) is the graph shift operator (GSO), denoted by $S$, which defines local interactions between neighboring nodes. Multiplying $S$ by a univariate signal $x \in \mathbb{R}^n$ (or $\mathbb{C}^n$) produces a new signal whose value at each node is a weighted sum of the signal values at its neighbors \citep{Sandryhaila2013discrete, Segarra2018}. Common choices for GSOs include the weighted adjacency matrix and the graph Laplacian. For an undirected graph, the Laplacian is defined as $L = \operatorname{diag}(\{\sum_{j=1}^n w_{ij}\}_{i=1}^{n}) - W$, where $w_{ij}$ is the $(i,j)$th element of $W$.  
	%
	%
	In general settings, the GSO $S$ is assumed to be normal, allowing it to be decomposed as $S = V \Lambda V^H$, where $V = (v_1 \mid \cdots \mid v_n)$ contains orthonormal eigenvectors and $\Lambda = \operatorname{diag}(\lambda_1, \ldots, \lambda_n)$ with $\lambda_1 \le \lambda_2 \le \cdots \le \lambda_n$, the eigenvalue $\lambda_i$ can be interpreted as the graph frequency corresponding to the basis vector $v_i$. Graph frequencies are ordered according to the total variation of the eigenvectors. Following \cite{Sandryhaila2014} and \cite{Singh2016graph}, eigenvectors with smaller total variation are considered low-frequency components, while those with larger variation are regarded as high-frequency components. In particular, when $S=L$, the graph frequencies are real-valued, and their ordering coincides with the usual numerical ordering.
	
	{\color{black}
		Given the eigendecomposition $S = V \Lambda V^H$ with eigenvalues ordered from low to high frequencies, the graph Fourier transform (GFT) of a signal $x$ is defined as $\tilde{x}=V^H x$.
		The $\ell$th element, $\tilde{x}(\lambda_\ell)=v_\ell^H x$, is called the graph Fourier coefficient of $x$ at graph frequency $\lambda_\ell$. The GFT therefore transforms a signal from the node domain into the graph frequency (or spectral) domain, similarly to the classical Fourier transform.}
	A linear and shift-invariant (LSI) graph filter is a linear operator that modifies graph spectral components in a frequency-dependent manner. It can be written as $\mr{H} = \sum_{\ell=1}^{n} h_\ell S^{\ell-1} = V\operatorname{diag}(\{ h(\lambda_\ell)\}_{\ell=1}^n)V^H$, where $h(\lambda) = \sum_{\ell=1}^{n} h_\ell \lambda^{\ell-1}$ denotes the frequency response of the filter \citep{Sandryhaila2013discrete}. The filtered output of $x$ is then $\mr{H}x$. These filters are essential in GSP for tasks, such as denoising, smoothing, and signal reconstruction, on graphs \citep{Tremblay2018design}. In this paper, the term `graph filter' specifically refers to LSI graph filters unless stated otherwise.
	
	\subsection{Stationarity and Spectral Densities on Graphs} \label{sec:graph_stationarity}
	The concept of weak stationarity for random processes on graphs has been widely used in statistical graph signal processing \citep{kim2025cross, Perraudin2017, Marques2017}. The definition of weak stationarity for multivariate random graph processes below was introduced by \citet{kim2024principal} as an extension of the univariate notion in \citet{Marques2017}. 
	
	\begin{definition} \label{def:multivariate_stationarity}
		\citep{kim2024principal} Given a normal graph shift operator $S$, consider a zero-mean $p$-dimensional random graph process $X=(X_1 \mid \cdots \mid X_p)$, where each $X_i$ ($1 \le i \le p$) is an $n$-dimensional vector. The random graph process $X$ is said to be weakly stationary if the cross-covariance matrices $\Sigma_{ij}^X:= \operatorname{Cov}(X_i, X_j)$ and $S$ are simultaneously diagonalizable for $1\le i, j \le p$.
	\end{definition}
	
	If $X_i$ ($1 \le i \le p)$ are not zero-mean random graph processes, their first moments are required to satisfy $\mathbb{E}[X_i] \propto v_j$ ($1 \le j \le n)$. 
	This definition is relative to the chosen GSO $S$, which serves as a signal shift on a graph. It implies that the covariance and cross-covariance matrices are diagonalized by the graph Fourier basis associated with $S$. This is analogous to the time-series case, in which the covariance structure of a weakly stationary process is diagonalized by the standard Fourier basis. Moreover, as discussed in \cite{kim2025cross}, \Cref{def:multivariate_stationarity} is equivalent to a graph-domain shift-invariance condition. This equivalence links graph stationarity to the familiar time-series concept of weak stationarity, in which second-order moments are invariant to time shifts and depend only on relative lags.
	
	Although there is currently no widely accepted formal statistical test for graph stationarity, diagnostic measures have been proposed to quantify how closely a graph process satisfies weak stationarity with respect to a given GSO \citep{Perraudin2017, Marques2017, kim2025cross}. These diagnostics are based on the degree of diagonal concentration of the estimated covariance or cross-covariance structure in the graph Fourier domain.

	For a weakly stationary random graph process $X$, \cite{kim2025cross} defined the graph cross-spectral density (GCSD) between $X_i$ and $X_j$ ($1 \le i,j \le p)$ as an $n \times 1$ vector,  
	$
	p_{ij}^X = (p_{ij}^X(\lambda_1), \ldots, p_{ij}^X(\lambda_n))^\top = \operatorname{diag}(V^H \Sigma_{ij}^X V).
	$ 
	GCSD characterizes the frequency-wise dependence between two graph signals. When $i=j$, this becomes the graph power spectral density (GPSD) of $X_i$ introduced in \cite{Marques2017}, which shows the distribution of signal energy across graph frequencies. Both GPSD and GCSD can be estimated using the windowed average graph periodogram and graph cross-periodogram, respectively, with random windows applicable in practical implementations \citep{kim2025cross}. The (squared) graph coherence between $X_i$ and $X_j$ ($1 \le i,j \le p)$ was subsequently defined in \citet{kim2025cross} as an $n\times1$ vector $c_{ij}^X = (c_{ij}^X(\lambda_1), \ldots, c_{ij}^X(\lambda_n))^\top$, where
	$$c_{ij}^X(\lambda_\ell) = \frac{\lvert p_{ij}^X(\lambda_\ell)\rvert^2}{p_{ii}^X(\lambda_\ell) \,p_{jj}^X(\lambda_\ell)}, \quad 1 \le i,j \le p,\; 1 \le \ell \le n.$$
	
	For brevity, `weak stationarity' will hereafter be called `stationarity,' and `(squared) graph coherence' as `graph coherence.' The term `graph signal' is used interchangeably to refer to either a random graph process or its realization, depending on the context.
	
	\subsection{Related Work}  \label{sec:relatedwork}
	
	Several studies have integrated graph structures into CCA. \cite{yuan2014graph} proposed graph-regularized multiset canonical correlation analysis (GrMCCA) for multiple datasets measured on the same observations across multiple classes. For each dataset, they construct $k$-nearest-neighbor within-class and between-class graphs from observation similarities. These graphs are incorporated into the CCA objective to maximize correlations between canonical variables while promoting within-class compactness and between-class separation.

	\cite{chen2018canonical} proposed graph-regularized CCA (gCCA), which assumes low-dimensional common latent sources generating the two datasets, aligning with the perspective of \cite{bach2005probabilistic}. 
	The latent sources are assumed to form a smooth multivariate graph signal $s=(s_1,\ldots,s_n)^\top$, where $s_i \in \mathbb{R}^\rho$, on a source graph $\mc{G}_s$, whose nodes represent the $n$ observations. The source graph may be given or constructed from prior knowledge. Since the canonical variables $\alpha_i$ and $\beta_j$ serve as one-dimensional approximations of $s_i$ and $s_j$, respectively, the canonical variable vectors $\alpha=(\alpha_1,\ldots, \alpha_n)^\top$ and $\beta=(\beta_1,\ldots, \beta_n)^\top$ are obtained by maximizing $\operatorname{Corr}(\alpha,\beta) - \gamma \alpha^\top L_s \beta$, where $L_s$ denotes the graph Laplacian of $\mc{G}_s$. \cite{chen2019graph} further extended gCCA to multiview settings, resulting in MGCCA. These methods operate in the node domain and impose smoothness constraints on the canonical variables. While this assumption is often valid, it does not always apply. Additionally, these methods do not explicitly address the interpretability of the resulting canonical variables: although the solutions are affected by graph regularization, it is difficult to understand them in terms of the underlying graph structure.

	In contrast, the proposed method incorporates graph structure in the frequency domain, providing a more general framework beyond cases where canonical variables on neighboring nodes are expected to have similar values. It produces interpretable canonical variables from a multiscale spectral perspective, offering insights unavailable in existing node-domain approaches. We compare the proposed method with gCCA by \cite{chen2018canonical} via an image classification task in Appendix~E, as it is the most relevant baseline.
	We further note that the recent graph CCA method of \cite{park2025graph} fundamentally differs from the previous methods discussed above and from our method because it uses a bipartite graph whose nodes represent the variables of the two datasets, rather than a graph structure based on observations.  
	
	\section{Proposed Method} \label{sec:proposed}
	\subsection{Graph Canonical Coherence Analysis and Its Properties} \label{subsec:cca_on_graph}
	Let $X = (X_1 \mid \cdots \mid X_p)$ and $Y = (Y_1 \mid \cdots \mid Y_q)$ denote $p$-dimensional and $q$-dimensional graph signals on a graph $\mathcal{G}$, respectively. 
	We construct an $r$-dimensional graph signal $Z = (Z_1 \mid \cdots \mid Z_r)$ for $r \le \min \,(p, q)$, where each component is defined as a sum of graph-filtered outputs of signals $X_1, \ldots, X_p$,
	\begin{equation} \label{eq:gcca_reduce}
		Z_i = \mr{H}_{i1}X_1 + \cdots + \mr{H}_{ip}X_p =: \mc{T}_i^{\,\mr{H}}(X), \quad 1 \le i \le r.
	\end{equation}
	Here, $\mr{H}_{ij}=V \operatorname{diag}(\{h_{ij}(\lambda_\ell)\}_{\ell=1}^{n}) V^H$ is a graph filter, and $\mc{T}_i^{\,(\cdot)}(\cdot)$ denotes a linear operator that sums graph-filtered outputs. 
	Similarly, we construct an $r$-dimensional graph signal $W = (W_1 \mid \cdots \mid W_r)$, whose components are defined as sums of graph-filtered versions of $Y_1, \ldots, Y_q$,
	\begin{equation} \label{eq:gcca_reduce_sym}
		W_i = \mr{F}_{i1}Y_1 + \cdots + \mr{F}_{iq}Y_q =:  \mc{T}_i^{\,\mr{F}}(Y), \quad 1 \le i \le r,
	\end{equation}
	where $\mr{F}_{ij} = V\operatorname{diag}(\{f_{ij}(\lambda_\ell)\})_{\ell=1}^n)V^H$ is a graph filter. We define $h_i(\lambda_\ell)=(h_{i1}(\lambda_\ell), \ldots, h_{ip}(\lambda_\ell))^H$ and $f_i(\lambda_\ell)=(f_{i1}(\lambda_\ell), \ldots, f_{iq}(\lambda_\ell))^H$.
	
	In a manner similar to classical CCA, we aim to sequentially find $Z_i$ and $W_i$ by maximizing the graph coherence between $Z_i$ and $W_i$ at each graph frequency, subject to the constraint that their graph power spectral densities are equal to 1 across all graph frequencies. Additionally, $Z_i$ and $W_i$ are constrained to have zero coherence with $Z_j$ and $W_j$ for $j=1, \ldots, i-1$ across all graph frequencies. The resulting $Z_i$ and $W_i$ are described in \Cref{thm:gcca_sym}.

	\begin{theorem} \label{thm:gcca_sym}
		Let $XY=(X_1 \mid \cdots \mid X_p \mid Y_1 \mid \cdots \mid Y_q)$ be a $(p+q)$-dimensional stationary graph signal with respect to the graph shift operator $S=V\Lambda V^H$ on $\mc{G}$. Let $p_{ij}^X$ and $p_{ij}^Y$ denote the graph cross-spectral densities of ($X_i, X_j$) and ($Y_i, Y_j$), respectively, and let $p_{ij}^{XY}$ and $p_{ij}^{YX}$ denote the graph cross-spectral densities of ($X_i, Y_j$) and ($Y_i, X_j$), respectively. For each graph frequency $\lambda_\ell$ ($1 \le \ell \le n$), define the spectral matrices, 
		$P_X(\lambda_\ell) \in \mathbb{C}^{p \times p},\;
		P_Y(\lambda_\ell) \in \mathbb{C}^{q \times q}, \; 
		P_{XY}(\lambda_\ell) \in \mathbb{C}^{p \times q}, \;
		P_{YX}(\lambda_\ell) \in \mathbb{C}^{q \times p}$,
		whose $(i,j)$th elements are $p_{ij}^X(\lambda_\ell)$, $p_{ij}^Y(\lambda_\ell)$, $p_{ij}^{XY}(\lambda_\ell)$, and $p_{ij}^{YX}(\lambda_\ell)$, respectively. Suppose that $P_X(\lambda_\ell)$ and $P_Y(\lambda_\ell)$ are nonsingular for all $1\le \ell \le n$. 
		
		The solutions maximizing the graph coherence of $Z_i$ and $W_i$ across each graph frequency, subject to the constraints that (i) the graph power spectral densities of $Z_i$ and $W_i$ are equal to 1 across all frequencies, and (ii) $Z_i$ and $W_i$ have zero coherence with $Z_j$ and $W_j$ for $j=1, \ldots, i-1$ across all graph frequencies, are given by
		\begin{eqnarray*}
			h_i(\lambda_\ell) &\propto& P_X(\lambda_\ell)^{-1} P_{XY}(\lambda_\ell)P_{Y}(\lambda_\ell)^{-1/2}\eta_i(\lambda_\ell), \quad 1 \le i \le r, \\
			f_i(\lambda_\ell) &\propto& P_Y(\lambda_\ell)^{-1/2} \eta_i(\lambda_\ell), \quad 1 \le i \le r,
		\end{eqnarray*}
		where $h_i(\lambda_\ell)$ and $f_i(\lambda_\ell)$ are normalized such that  $h_i(\lambda_\ell)^H P_X(\lambda_\ell) h_i(\lambda_\ell)= 1$ and $f_i(\lambda_\ell)^H P_Y(\lambda_\ell) f_i(\lambda_\ell)= 1$, thereby satisfying  constraint (i).
		Here,  $\eta_i(\lambda_\ell)$ is the $i$th orthonormal eigenvector of $P_Y(\lambda_\ell)^{-1/2} P_{YX}(\lambda_\ell)P_{X}(\lambda_\ell)^{-1}P_{XY}(\lambda_\ell)P_{Y}(\lambda_\ell)^{-1/2}$ corresponding to the $i$th largest eigenvalue $\gamma_i(\lambda_\ell)$. They satisfy  
		\begin{equation*}
			\begin{gathered}
				P_X(\lambda_\ell)^{-1} P_{XY}(\lambda_\ell)P_{Y}(\lambda_\ell)^{-1}P_{YX}(\lambda_\ell) h_i(\lambda_\ell) = \gamma_i(\lambda_\ell)h_i(\lambda_\ell),\\
				P_Y(\lambda_\ell)^{-1} P_{YX}(\lambda_\ell)P_{X}(\lambda_\ell)^{-1}P_{XY}(\lambda_\ell) f_i(\lambda_\ell) = \gamma_i(\lambda_\ell)f_i(\lambda_\ell).
			\end{gathered}
		\end{equation*}
		The maximum coherence between $Z_i$ and $W_i$ is $\gamma_i(\lambda_\ell)$. Moreover, $Z_i$ and $W_i$ have zero coherence with $Z_j$ and $W_j$ for $j < i$.
	\end{theorem}
	A proof is provided in Appendix~B.
	The pair of $Z_i$ and $W_i$ in \Cref{thm:gcca_sym} is called the \textit{$i$th pair of (population) canonical graph signals}. Their graph coherence, $\gamma_i(\lambda_\ell)$, is called the \textit{$i$th (population) graph canonical coherence}. The entire procedure for implementing CCA on graphs is referred to as `gCChA'. This procedure can alternatively be formulated from an error minimization perspective, as described in Appendix~A.

	To implement gCChA, the detailed procedure is outlined in Algorithm C.1 of Appendix~C, which provides sample versions of the canonical graph signal pairs and graph canonical coherences, denoted by $\{(\hat{Z}_i, \hat{W}_i)\}_{i=1}^r$ and $\hat{\gamma}_i(\lambda_\ell)$. The proposed method is implemented by repeating the procedure for the population canonical graph signals, replacing the population GCSDs with their estimates, since the true GCSDs are unknown. For this, one can use either the graph cross-periodogram or the windowed average graph cross-periodogram proposed by \cite{kim2025cross}. If multiple realizations of the two sets of graph signals $X$ and $Y$ are available, the graph cross-periodogram is used. If only a small number of realizations are available, the windowed average graph cross-periodogram with random windows is used.

	\subsection{Interpretation of Graph Canonical Coherence Analysis} \label{subsec:interpretation}
	In classical CCA, interpretation is typically based on correlations between the original variables and their respective canonical variates, known as canonical loadings, rather than on the canonical coefficients themselves, which can be unstable in the presence of multicollinearity. Furthermore, correlations between the original variables and the canonical variates of the opposite set, called canonical cross-loadings, can be used to express the relationship between a variable in one set and a canonical variate in the other. In gCChA, we similarly compute the graph coherences between the original graph signals in each dimension and the respective or opposite canonical graph signals, called \textit{graph canonical loadings} and \textit{graph canonical cross-loadings}, respectively. These quantities facilitate interpretation in the graph frequency domain, representing the proportion of variance in the graph Fourier coefficients of each variable that is linearly shared with the canonical graph signals from either the same set or the opposite set. Because the graph coherences contain information at each graph frequency $\lambda_\ell$, we can provide a multiscale interpretation across graph frequencies. For example, at a low graph frequency, a high graph coherence value between a canonical graph signal and a particular original signal indicates that they are closely related at a coarse scale. Conversely, at high graph frequency, a high graph coherence value indicates a close relationship at fine scales. An example of a more detailed interpretation is presented in \Cref{sec:g20}.
	
	Motivated by well-established terminology in classical CCA, we introduce similar quantities in the gCChA framework. The \textit{graph canonical communality}, defined as the sum of graph canonical loadings for each canonical graph signal, indicates how much of the variance in each variable’s graph Fourier coefficients can be explained by the canonical graph signals. In other words, it reflects the importance of each variable in establishing the canonical relationship between the two sets within the graph frequency domain. Specifically, the graph canonical communality of $X_j$ at graph frequency $\lambda_\ell$, is given by $\sum_{i=1}^r c_{ij}^{ZX}(\lambda_\ell)$, and that of $Y_j$ is $\sum_{i=1}^r c_{ij}^{WY}(\lambda_\ell)$, where $c_{ij}^{ZX}(\lambda_\ell)$ and $c_{ij}^{WY}(\lambda_\ell)$ represent the coherences between $Z_i$ and $X_j$ and between $W_i$ and $Y_j$, respectively, at graph frequency $\lambda_\ell$, respectively, defined by 
	\[
	c_{ij}^{ZX}(\lambda_\ell) = \frac{\lvert p_{ij}^{ZX}(\lambda_\ell) \rvert^2}{p_{ii}^Z(\lambda_\ell) p_{jj}^X(\lambda_\ell)}~~~\text{and}~~~ c_{ij}^{WY}(\lambda_\ell) = \frac{\lvert p_{ij}^{WY}(\lambda_\ell) \rvert^2}{p_{ii}^W(\lambda_\ell) p_{jj}^Y(\lambda_\ell)}.
	\]
	
	To evaluate the explanatory power of the canonical graph signals at each graph frequency $\lambda_\ell$, we define the \textit{graph canonical adequacy} as the average of all graph canonical loadings for the original graph signals in a given set with respect to a single canonical graph signal. This quantity reflects how well the canonical graph signals summarize the variability of each set of graph signals at each frequency in the graph frequency domain. Specifically, the graph canonical adequacy of $Z_i$ at graph frequency $\lambda_\ell$ is given by $\sum_{j=1}^p c_{ij}^{ZX}(\lambda_\ell)/p$, and that of $W_i$ is $\sum_{j=1}^q c_{ij}^{WY}(\lambda_\ell)/q$. As shown in \Cref{prop:explainpower}, the graph canonical communalities range from 0 to 1, and the cumulative explanatory power, obtained by summing the graph canonical adequacies up to each component, also falls between 0 and 1. At a given graph frequency $\lambda_\ell$, the cumulative explanatory power of the first $t$ ($t\le r$) canonical graph signals $Z_1,\ldots, Z_t$, defined as $\sum_{i=1}^t \sum_{j=1}^p c_{ij}^{ZX}(\lambda_\ell)/p$, represents the proportion of the total variance explained by these canonical graph signals at frequency $\lambda_\ell$ in the graph frequency domain. Similarly, the cumulative explanatory power of the first $t$ canonical graph signals $W_1,\ldots, W_t$, given by $\sum_{i=1}^t \sum_{j=1}^q c_{ij}^{WY}(\lambda_\ell)/q$, quantifies the proportion of variance explained by these canonical graph signals at frequency $\lambda_\ell$.
	
	\begin{proposition} \label{prop:explainpower}
		For $1 \le \ell \le n$, the following inequalities hold: 
		$
		0 \le \sum_{i=1}^r c_{ij}^{ZX}(\lambda_\ell) \le 1,~ j=1,\ldots, p, \mbox{ and }~
		0\le \sum_{i=1}^r c_{ij}^{WY}(\lambda_\ell) \le 1,~ j=1,\ldots, q. 
		$
		Consequently, 
		$
		0 \le \frac{1}{p}\sum_{i=1}^r \sum_{j=1}^p c_{ij}^{ZX}(\lambda_\ell) \le 1,\mbox{ and  }~ 0 \le \frac{1}{q}\sum_{i=1}^r \sum_{j=1}^q c_{ij}^{WY}(\lambda_\ell) \le 1$.  
		Furthermore, if $r=\min(p,q)$, then
		$
		\sum_{i=1}^r c_{ij}^{ZX}(\lambda_\ell)=1~ \text{when } p<q, \mbox{ and }
		\sum_{i=1}^r c_{ij}^{WY}(\lambda_\ell)=1~ \text{when } p>q.
		$	
	\end{proposition}
	A proof is provided in Appendix~B.
	%
	We note that both the interpretations and the proposition in this section directly apply to the sample versions.
	
	\section{Numerical Experiments} \label{sec:realdata}
	{\color{black}
		\subsection{Simulation Study}
		
		We conduct a simulation study to evaluate whether the proposed gCChA method can recover population graph canonical coherences in a setting where the underlying graph and the target graph-frequency quantities are fixed and explicitly defined.
		
		We generate a connected random geometric graph with $n=50$ nodes. The nodes are independently sampled from the unit square, and two nodes are connected when their Euclidean distance is smaller than a prespecified radius. The weight of an edge between nodes $u$ and $v$ is defined by $w_{uv}
		=
		\exp\left\{d^2(u,v) / \bar d^2
		\right\}$,
		where $d(u,v)$ is the Euclidean distance between the two nodes and $\bar d$ is the average pairwise distance. The graph Laplacian is used as the graph shift operator, whose eigenvalues are denoted by $\lambda_1 \le \cdots \le \lambda_n$. The graph and its associated graph Fourier basis are kept fixed throughout the simulation.
		
		We consider two bivariate graph signals, 
		$X=(X_1 \mid X_2)$ and
		$Y=(Y_1 \mid Y_2)$. Let $P_X(\lambda_\ell)$, $P_Y(\lambda_\ell)$, and $P_{XY}(\lambda_\ell)$ denote the population spectral matrices of $X$, $Y$, and their cross-spectral matrix, respectively, at graph frequency $\lambda_\ell$. We examine two scenarios: (i) diagonal within-set spectral matrices, in which $P_X(\lambda_\ell)$ and $P_Y(\lambda_\ell)$ are diagonal, and (ii) correlated within-set spectral matrices, in which $P_X(\lambda_\ell)$ and $P_Y(\lambda_\ell)$ are non-diagonal.
		In both scenarios, we specify smoothly varying population graph canonical coherences as
		\begin{equation*}
			\gamma_i(\lambda_\ell)     =
			\rho_i^2(\lambda_\ell),
			\quad i=1,2,
		\end{equation*}
		where $\rho_1(\lambda_\ell) =0.90-
		0.25\left\{
		1-\exp(-2\tilde{\lambda}_\ell)
		\right\}$ and $\rho_2(\lambda_\ell) = 0.25+
		0.20\left\{
		1-\exp(-2\tilde{\lambda}_\ell)
		\right\}$, with $\tilde\lambda_\ell =(\lambda_\ell-\lambda_1)/(\lambda_n-\lambda_1)$. By construction, $\rho_1(\lambda_\ell)>\rho_2(\lambda_\ell)$
		for all graph frequencies.
		
		Specifically, in Scenario 1, we set  
		\begin{equation*}
			P_X(\lambda_\ell)=I_2,
			\quad
			P_Y(\lambda_\ell)=I_2, \quad P_{XY}(\lambda_\ell)
			=
			\begin{pmatrix}
				\rho_1(\lambda_\ell) & 0\\
				0 & \rho_2(\lambda_\ell)
			\end{pmatrix},   
		\end{equation*}
		where $I_2$ denotes the $2 \times 2$ identity matrix. It follows from \Cref{thm:gcca_sym} that the corresponding population graph canonical coherences satisfy $\gamma_i(\lambda_\ell)  = \rho_i^2(\lambda_\ell)$. In Scenario 2, we introduce within-set spectral dependence by setting
		\begin{equation*}
			P_X(\lambda_\ell) = \begin{pmatrix}
				1 & \theta_X(\lambda_\ell)\\
				\theta_X(\lambda_\ell) & 1
			\end{pmatrix},
			\quad
			P_Y(\lambda_\ell) = \begin{pmatrix}
				1 & \theta_Y(\lambda_\ell)\\
				\theta_Y(\lambda_\ell) & 1
			\end{pmatrix},
		\end{equation*}
		where $\theta_X(\lambda_\ell)
		=
		0.35\exp(-1.5\tilde\lambda_\ell)$ and 
		$\theta_Y(\lambda_\ell)
		=
		0.25
		+
		0.15\left\{
		1-\exp(-2\tilde\lambda_\ell)
		\right\}$.
		The cross-spectral matrix is constructed as
		\begin{equation*}
			P_{XY}(\lambda_\ell)
			=
			P_X(\lambda_\ell)^{1/2}
			U_1
			D(\lambda_\ell)
			U_2^\top
			P_Y(\lambda_\ell)^{1/2},    
		\end{equation*}
		where $D(\lambda_\ell)
		=
		\operatorname{diag}
		\left\{
		\rho_1(\lambda_\ell),
		\rho_2(\lambda_\ell)
		\right\}$, and $U_1$ and $U_2$ are rotation matrices with rotation angles $\pi/6$ and $\pi/5$, respectively. By \Cref{thm:gcca_sym}, this construction yields the population graph canonical coherences $\gamma_i(\lambda_\ell)  = \rho_i^2(\lambda_\ell)$. 
		
		Under weak stationarity, the variances and covariances of the graph Fourier coefficients at frequency $\lambda_\ell$ are given by the corresponding graph power and cross-spectral densities defined in \Cref{sec:graph_stationarity}. Accordingly, the joint covariance structure of the graph Fourier coefficients can be specified using $P_X(\lambda_\ell)$, $P_Y(\lambda_\ell)$, and $P_{XY}(\lambda_\ell)$.
		For each graph frequency $\lambda_\ell$, we generate $R$ independent realizations in the graph frequency domain according to
		\begin{equation*}
			\begin{pmatrix}
				\tilde X^{(r)}(\lambda_\ell)\\
				\tilde Y^{(r)}(\lambda_\ell)
			\end{pmatrix}
			\sim
			N\left(
			0,
			\begin{pmatrix}
				P_X(\lambda_\ell) & P_{XY}(\lambda_\ell)\\
				P_{YX}(\lambda_\ell) & P_Y(\lambda_\ell)
			\end{pmatrix}
			\right), \quad r=1,\ldots,R,
		\end{equation*}
		where $\tilde X^{(r)}(\lambda_\ell) = (\tilde{X}^{(r)}_1(\lambda_\ell), \tilde{X}^{(r)}_2(\lambda_\ell))^\top$ and $\tilde Y^{(r)}(\lambda_\ell) = (\tilde{Y}^{(r)}_1(\lambda_\ell), \tilde{Y}^{(r)}_2(\lambda_\ell))^\top$,
		whose each entries represent the graph Fourier coefficients of the corresponding components at graph frequency $\lambda_\ell$.
		The graph Fourier coefficient vectors are generated independently across distinct graph frequencies and then transformed to the node domain using the inverse graph Fourier transform, yielding
		\begin{equation*}
			X^{(r)} = (X^{(r)}_1 \mid X^{(r)}_2), \quad Y^{(r)} = (Y^{(r)}_1 \mid Y^{(r)}_2), \quad  r=1,\ldots,R.
		\end{equation*}
		By construction, the resulting graph signals satisfy the weak stationarity condition in \Cref{def:multivariate_stationarity}.

		We consider $R\in\{50,100,200,400\}$. For each value of $R$, the simulation is repeated 100 times. In each repetition, the estimation accuracy of the $i$th graph canonical coherence is evaluated using the mean squared error (MSE), defined as $\sum_{\ell=1}^n
		\left\{
		\hat\gamma_i(\lambda_\ell)
		-
		\gamma_i(\lambda_\ell)
		\right\}^2 / n$, where $\hat\gamma_i(\lambda_\ell)$ denotes the estimated graph canonical coherence.
		
		\Cref{fig:gccha_scenario} compares the estimated graph canonical coherences with their population counterparts for a representative simulation run under the two simulation scenarios. The curves are shown based on $R=200$ realizations. In both scenarios, the estimated curves closely follow the overall frequency-dependent patterns of the population graph canonical coherences. The first graph canonical coherence decreases from low to high graph frequencies, whereas the second increases, and these contrasting patterns are successfully recovered by the proposed method.
		
		\begin{figure}[!t]
			\centering
			\includegraphics[width=0.67\linewidth]{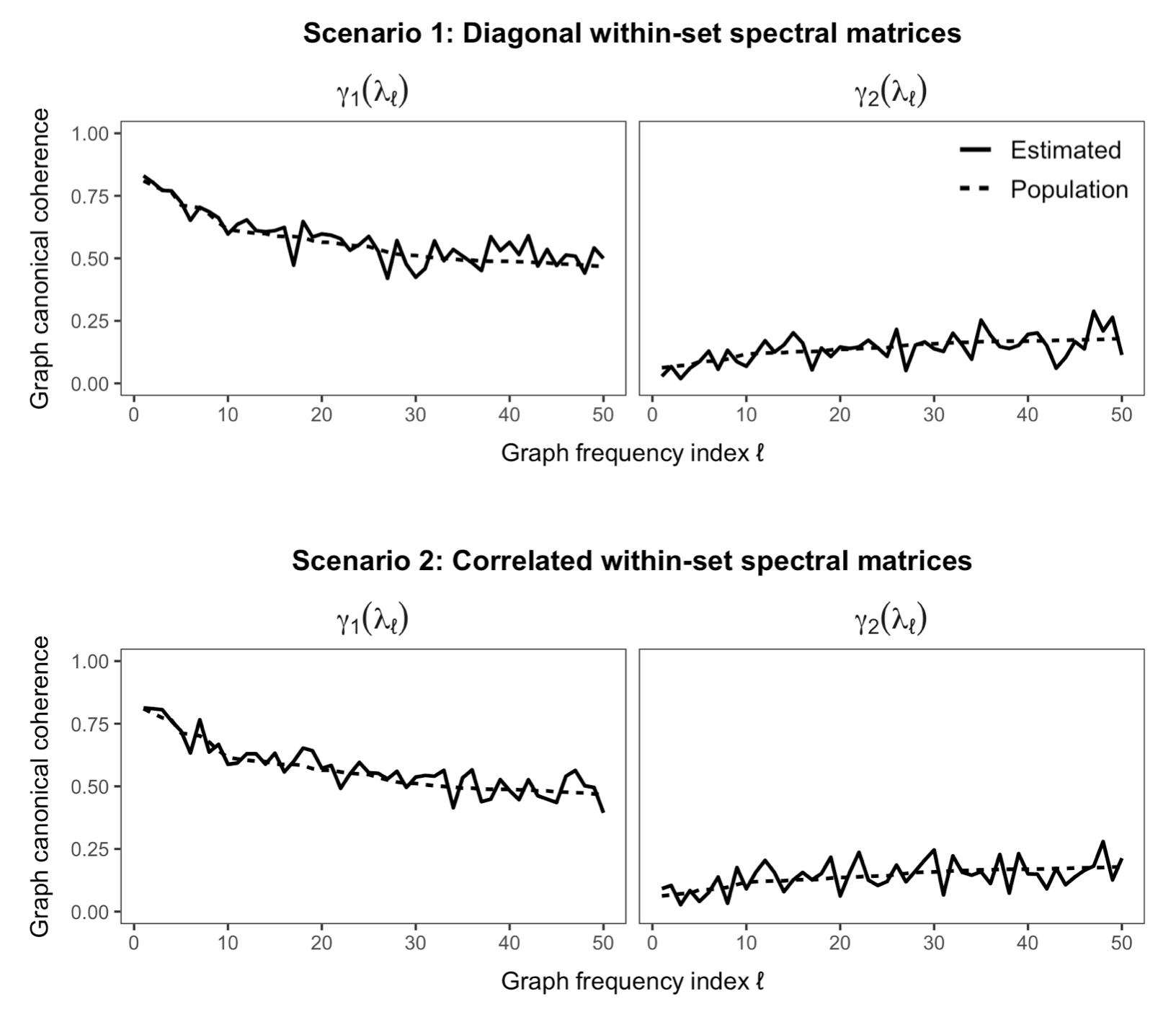}
			\vspace{-3mm}
			\caption{
				Estimated graph canonical coherence curves and their population counterparts under the two simulation scenarios.
			}
			\label{fig:gccha_scenario}
		\end{figure}
		
		\begin{figure}[!t]
			\centering
			\includegraphics[width=0.67\linewidth]{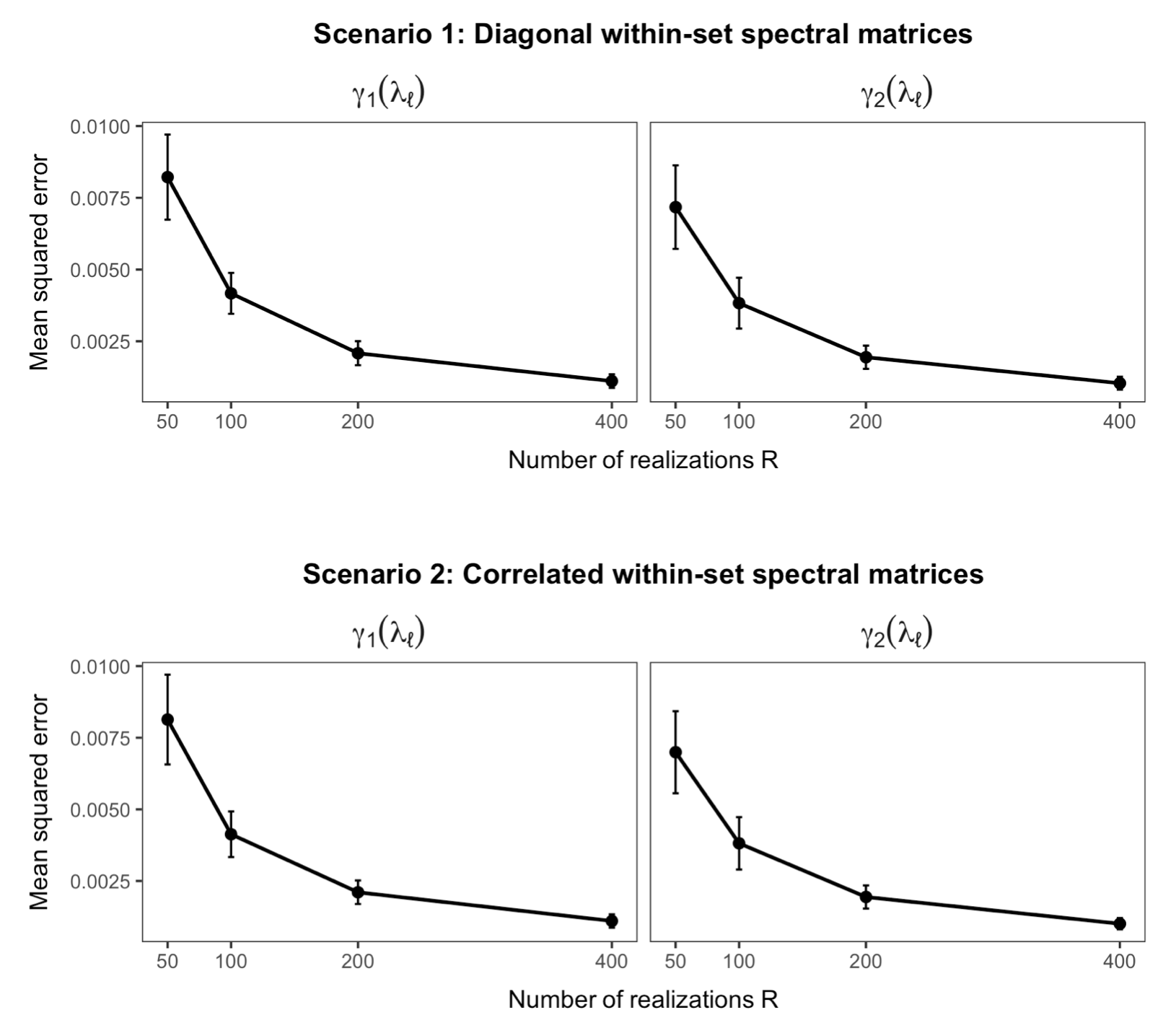}
			\vspace{-3mm}
			\caption{
				Mean squared error of the estimated graph canonical coherences under the two simulation scenarios. The points and error bars represent the averages and standard deviations over 100 Monte Carlo replications, respectively.
			}
			\label{fig:gccha_scenario_graphs}
		\end{figure}

		\Cref{fig:gccha_scenario_graphs} reports the average MSE values over 100 Monte Carlo replications, with error bars representing one Monte Carlo standard deviation. For both simulation settings and both canonical components, the MSE decreases steadily as the number of realizations $R$ increases. The results are also similar across the two scenarios, indicating that the estimation accuracy is not substantially affected by within-set spectral correlations. The Monte Carlo standard deviations likewise decrease as $R$ increases. Overall, these results demonstrate that gCChA accurately recovers the population graph canonical coherences on a fixed underlying graph and that the estimation accuracy improves as more realizations become available.
	}
	
	\subsection{G20 Network Data Analysis} \label{sec:g20}
	We apply gCChA to data from the G20 network. We construct a graph with 18 nodes representing G20 countries, excluding Argentina and the European Union due to missing data. The edge weights are defined as $\log(1+\text{trade asset})$ based on trade data from 2019 \citep{kim2025gfa}. The resulting trading network is shown in \Cref{fig:trading_nw}, and the country codes are listed in Table D.1 of Appendix~D.
	\begin{figure}[!t]
		\centering
		\includegraphics[width=0.43\linewidth]{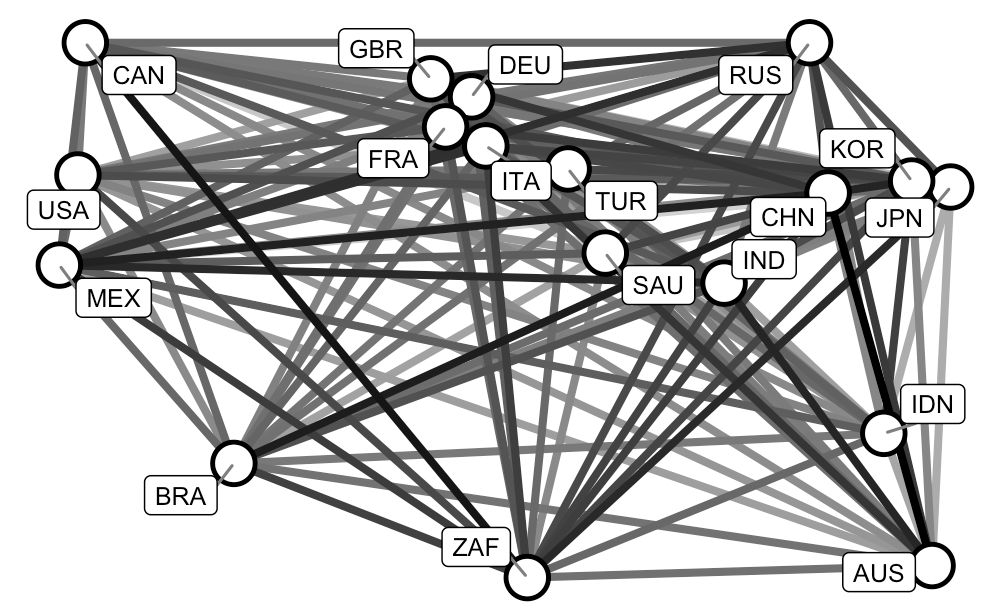}
		\vspace{-3mm}
		\caption{Trading network among G20 countries in 2019, where darker edges represent stronger trade connections.}
		\label{fig:trading_nw}
	\end{figure}
	%
	%
	Two sets of observed graph signals, $X$ and $Y$, denote economic and energy-related indicators, respectively. Each set consists of five graph signals, denoted as $X=(X_1 \mid \cdots \mid X_5)$ and $Y=(Y_1 \mid \cdots \mid Y_5)$, which correspond to (exports, foreign direct investment, GDP growth, imports, inflation) and (energy imports, energy use, fossil fuel consumption, GDP per unit of energy use, renewable energy consumption). Brief descriptions of the graph signals are provided in Table D.2 of Appendix~D, with detailed definitions available in the World Bank Group’s metadata glossary. The data are from the World Bank Open Data website (\url{https://data.worldbank.org/indicator}). In summary, we have $n=18$, $p=5$, and $q=5$.

\begin{figure}[!t]
		\centering
		\includegraphics[width=0.67\linewidth]{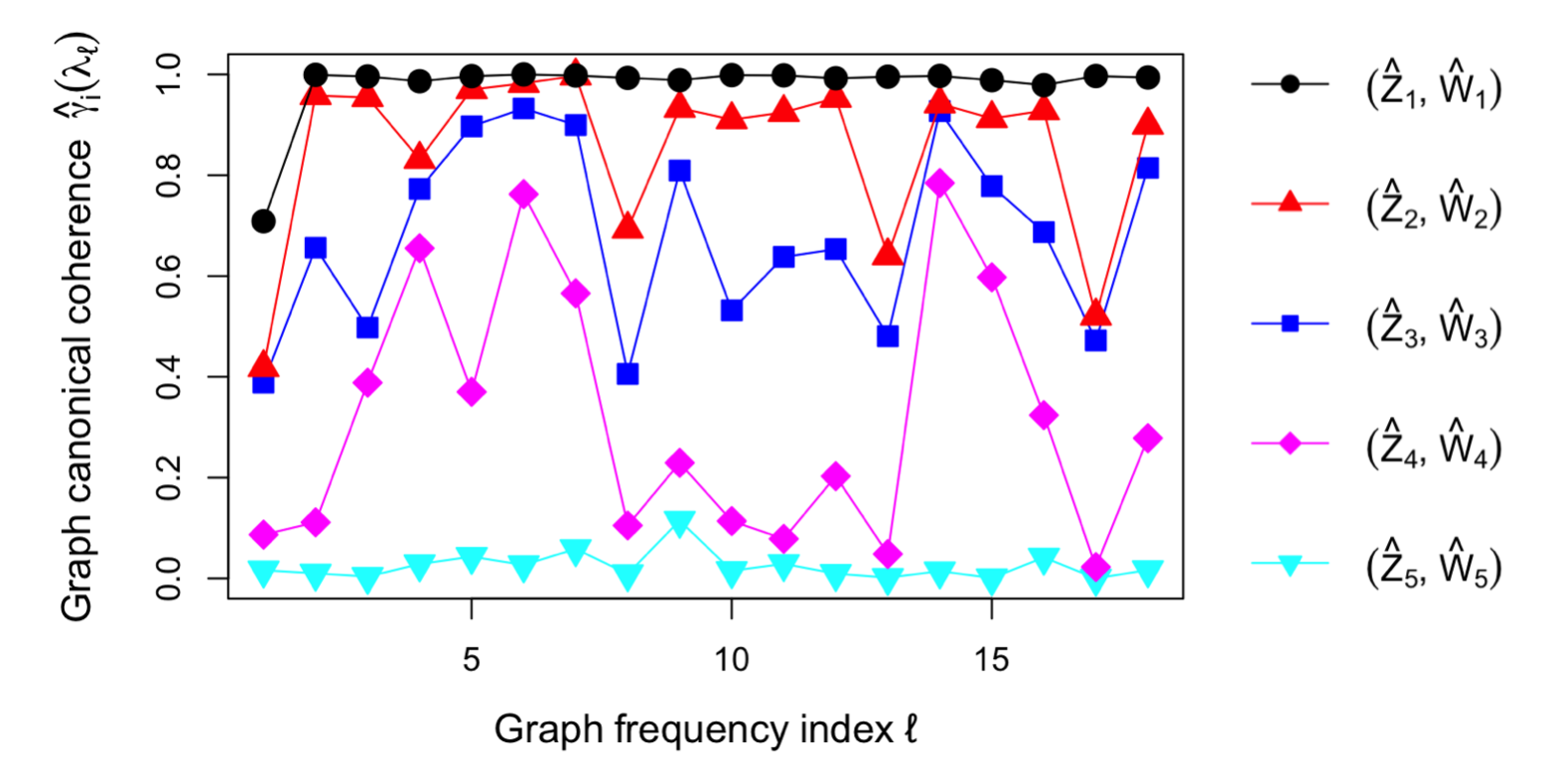}
		\vspace{-3mm}
		\caption{Graph canonical coherences $\hat{\gamma}_i(\lambda_\ell)$ for the canonical graph signal pairs $(\hat{Z}_i, \hat{W}_i)$, $i=1,\ldots,5$, across graph frequencies.}
		\label{fig:graph_canonical_coherence}
	\end{figure}

	In implementing gCChA, we use the graph Laplacian $L$ as GSO. The estimation of the graph power and cross-spectral densities is performed using the windowed averaging method with 50 random windows \citep{kim2025cross}. There are five pairs of canonical graph signals (i.e., $r=\min(p,q)=5$), $(\hat{Z}_1, \hat{W}_1), \ldots, (\hat{Z}_5, \hat{W}_5)$. The graph canonical coherences within each pair, $\hat{\gamma}_1(\lambda_\ell), \ldots, \hat{\gamma}_5(\lambda_\ell)$, are shown in \Cref{fig:graph_canonical_coherence}, where $\lambda_\ell$ denotes the graph frequencies of the trading network. As expected, the canonical coherences decrease from $(\hat{Z}_1, \hat{W}_1)$ to $(\hat{Z}_5, \hat{W}_5)$ across all graph frequencies. 
	
\begin{figure}[!t]
		\centering
		\includegraphics[width=0.9\textwidth]{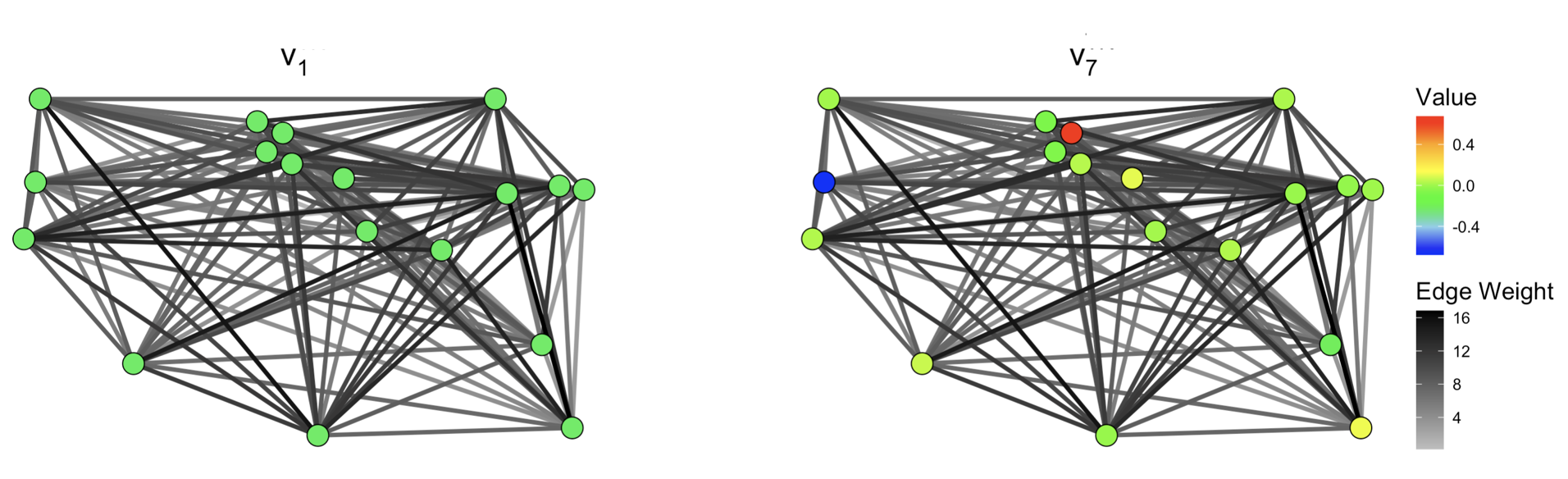}
		\vspace{-3mm}
		\caption{Basis signals $v_1$ and $v_7$.}
		\label{fig:tn_basis_signals}
	\end{figure}

	From \Cref{fig:graph_canonical_coherence}, we observe a noticeable gap between the graph canonical coherences of $(\hat{Z}_2, \hat{W}_2)$ and $(\hat{Z}_3, \hat{W}_3)$. We calculate the cumulative explanatory powers of the first two canonical graph signals, $\sum_{i=1}^2 \sum_{j=1}^p c_{ij}^{\hat{Z}\hat{X}}(\lambda_\ell)/p$ and $\sum_{i=1}^2 \sum_{j=1}^q c_{ij}^{\hat{W}\hat{Y}}(\lambda_\ell)/q$, which exceed 0.75, indicating that more than 75\% of the variance within $X$ and $Y$, on average across graph frequencies, is explained by the first two respective canonical graph signals. We therefore interpret the first two canonical pairs. Although interpretation can be applied at each graph frequency, we examine $\lambda_1$ and $\lambda_7$ as representative frequencies for coarse and fine scales, respectively. \Cref{fig:tn_basis_signals} shows the associated basis signals $v_1$ and $v_7$. The basis signal $v_1$ is constant across all nodes, representing the coarsest-scale pattern of the G20 network. In contrast, $v_7$ highlights the contrast between Germany and the United States.
	
	\begin{figure}[!t]
		\centering
		\includegraphics[width=0.9\textwidth]{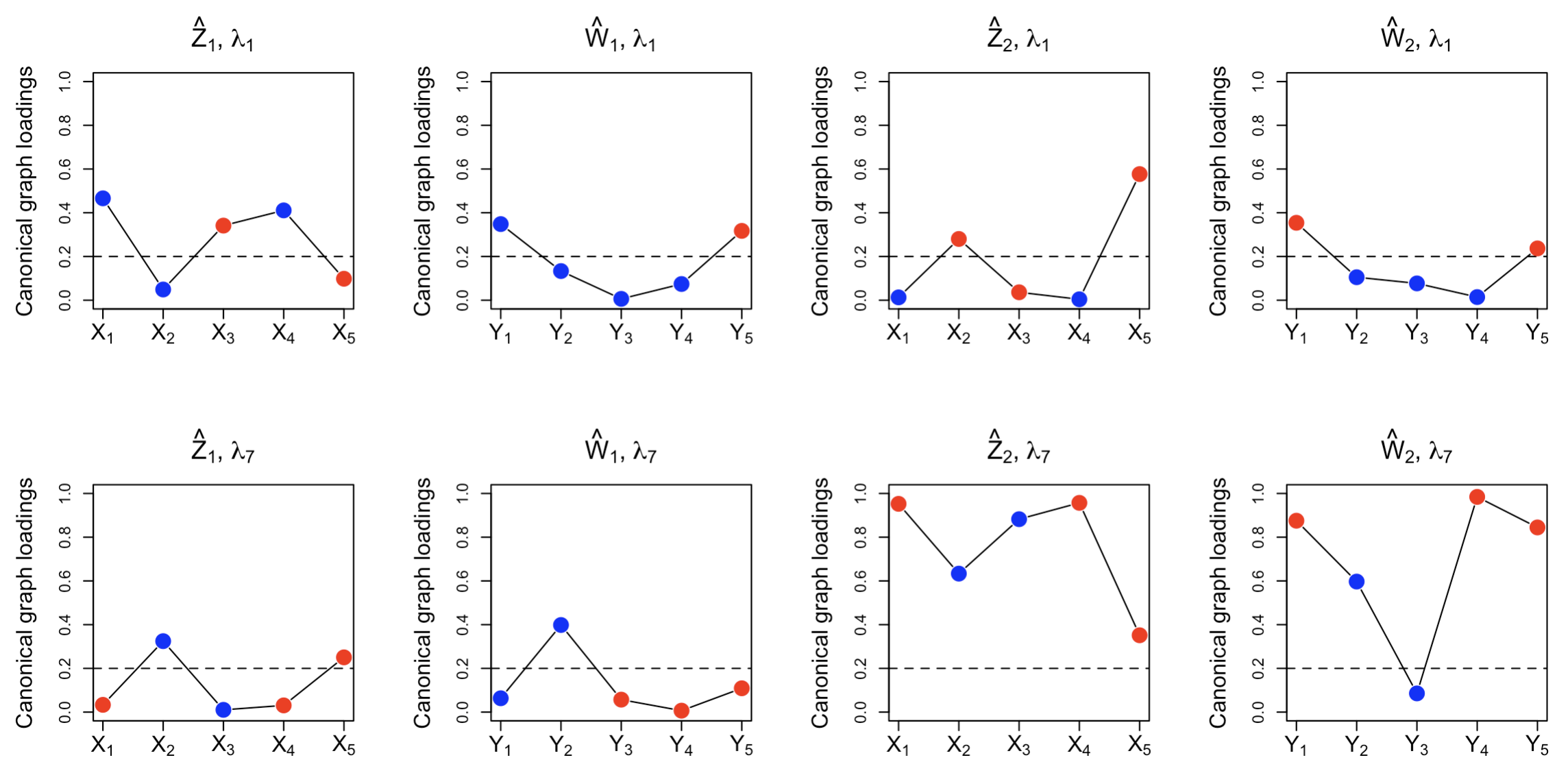}
		\vspace{-3mm}
		\caption{Graph canonical loadings of the first two canonical graph signals at graph frequencies $\lambda_1$ (top row) and $\lambda_7$ (bottom row). Red dots show positive correlations in the frequency domain, while blue dots represent negative correlations. Loadings with a magnitude exceeding 0.2 (shown by the dashed line) are used for interpretation.}
		\label{fig:graph_canonical_loadings}
	\end{figure}
	
	At the lowest frequency $\lambda_1$, the canonical graph signals depict broad relationships across the G20 network, capturing country-level similarity patterns that dominate at a global scale, as reflected by the constant basis signal $v_1$. At this frequency, as shown in the top row of \Cref{fig:graph_canonical_loadings}, $\hat{Z}_1$ is associated with $X_1$, $X_3$, and $X_4$ in the $(-),(+),(-)$ directions, while $\hat{W}_1$ is associated with $Y_1$ and $Y_5$ in the $(-)$ and $(+)$ directions, respectively. Therefore, the first canonical pair $(\hat{Z}_1, \hat{W}_1)$ shows that countries with higher GDP growth but lower trade volumes tend to have lower energy import dependence and greater renewable energy use. Countries such as France, the United Kingdom, Canada, Australia, and Brazil exhibit this pattern, reflecting growth- and sustainability-focused systems characterized by moderate trade activity, steady growth, and a relatively high level of energy independence supported by renewables, which tends to increase both $\hat{Z}_1$ and $\hat{W}_1$. In contrast, export- and industry-driven countries like Korea, Japan, Italy, India, and T\"{u}rkiye show the opposite trend, causing both $\hat{Z}_1$ and $\hat{W}_1$ to decrease. This reflects greater trade dependence and higher reliance on energy imports, and lower adoption of renewable energy. This canonical relationship  captures a broad, global-scale trade-energy connection across the G20 network, as the correlation between $\hat{Z}_1$ and $\hat{W}_1$ at frequency $\lambda_1$ reflects a consistent structural pattern shared by many countries. 
	For the second canonical pair, $\hat{Z}_2$ is associated with $X_2$ and $X_5$, and $\hat{W}_2$ with $Y_1$ and $Y_5$, all in the $(+)$ direction. This pair indicates that countries with higher levels of foreign direct investment (FDI) and inflation tend to have greater energy imports and increased use of renewable energy. This reflects a pattern in which economies with greater openness and inflationary pressures simultaneously increase energy imports and pursue energy transition policies. Countries such as China and India, known for rapid industrial growth, substantial FDI inflows, and ongoing investments in renewable energy, tend to push both $\hat{Z}_2$ and $\hat{W}_2$ higher. In contrast, Canada, Australia, and Saudi Arabia, which are resource-based and domestically stable economies with low inflation and limited reliance on imported energy, tend to pull these variables lower. Overall, the coarse-scale canonical signals reveal the main relationships among economic openness, price trends, and the energy transition observed across the G20 network.
	
	At the high frequency $\lambda_7$, the canonical graph signals highlight fine-scale country contrasts within the G20 trade–energy network, particularly between Germany and the United States, as illustrated by the corresponding basis signal $v_7$. For the first canonical pair, $\hat{Z}_1$ is associated with $X_2$ and $X_5$ in the $(-)$ and $(+)$ directions, while $\hat{W}_1$ is associated with $Y_2$ in the $(-)$ direction. This relationship indicates that countries with lower FDI inflows and total energy consumption tend to have higher values of $(\hat{Z}_1, \hat{W}_1)$. Germany, known for relatively modest FDI inflows but efficient and moderate energy use, tends to have higher values of $\hat{Z}_1$ and $\hat{W}_1$. In contrast, the United States, with higher FDI inflows and energy consumption, tends to have smaller values of $\hat{Z}_1$ and $\hat{W}_1$.
	The second canonical pair exhibits a more complex structure: $\hat{Z}_2$ is associated with $X_1, X_2, X_3, X_4,$ and $X_5$ in the $(+), (-), (-), (+), (+)$ directions, while $\hat{W}_2$ is connected to $Y_1, Y_2, Y_4,$ and $Y_5$ in the $(+), (-), (+), (+)$ directions, respectively. Unlike the first pair, which mainly shows the connection between capital inflows and energy use, this pair illustrates interconnected relationships among trade dependence, economic growth, and renewable energy strategies. Germany, known for its high trade openness, moderate GDP growth, reliance on imported energy, and strong energy efficiency, tends to have larger values of $\hat{Z}_2$ and $\hat{W}_2$. In contrast, the United States, often described as having lower trade dependence, higher GDP growth, and less reliance on energy imports, tends to have lower values of $\hat{Z}_2$ and $\hat{W}_2$. Overall, the high-frequency canonical signals at $\lambda_7$ emphasize localized variations distinguishing advanced economies with differing structural balances among industrial openness, energy efficiency, and policy direction.

	\section{Concluding Remarks} \label{sec:conclude}
	In this paper, we propose a graph canonical coherence analysis (gCChA) method that operates in the graph frequency domain. The proposed method handles two multivariate graph signals defined on the same graph and produces two canonical graph signals that serve as low-dimensional representations summarizing the dependence structure between the two and capturing their shared information. The proposed framework offers a spectral interpretation of canonical graph signals by associating them with the original graph signals at each graph frequency. 
	{\color{black}
		We demonstrated its applicability through a simulation study and an analysis of the G20 economic-energy network. The simulation results showed that gCChA accurately recovers the population graph canonical coherences under a fixed underlying graph and that its estimation accuracy improves as the number of graph-signal realizations increases. The real-data analysis further showed that the canonical graph signals enable meaningful spectral interpretation across different graph frequencies.}
	
	There are several directions for future work. One avenue is to extend the current framework to accommodate more than two graph datasets, enabling canonical dependence analysis among multiple graph signals. Another direction is to explore different applications with multiple graph-structured datasets, such as brain functional networks and environmental sensing networks. Further methodological development is also of interest, including improved GCSD estimation and optimal window design for the windowed-average graph cross-periodogram.
	Moreover, although spectral-domain analysis is central to interpreting the proposed gCChA framework, it requires diagonalizing the GSO, which can be computationally demanding for large-scale graphs. Fast and approximate graph Fourier transform techniques \citep{le2016flexible, le2017approximate, lu2019fast} may provide useful directions for reducing the computational burden in large-scale applications.
	
	A further issue concerns the sensitivity of graph spectral quantities to the choice of GSO. In graph signal processing, the graph and GSO are typically treated as part of the pre-specified data domain, so spectral analysis is interpreted relative to this choice. Accordingly, gCChA framework should be understood conditionally on the specified graph structure and GSO. Since different graph constructions or edge-weight perturbations may affect the graph Fourier basis, the resulting graph spectral density matrices and canonical coherences may also change. Developing perturbation analyses and sensitivity assessments for alternative GSO choices would be valuable future work, with studies on graph signal processing under graph perturbations \citep{ceci2020graph} serving as a useful starting point.
	Finally, relaxing the weak stationarity assumption is an important extension. In particular, locally stationary graph processes \citep{canbolat2024locally} may provide a useful foundation for modeling graph signals with node-varying spectral quantities. Developing a multivariate locally stationary graph process framework, along with estimation procedures for node-varying spectral quantities, would broaden the applicability of gCChA to heterogeneous graph-structured data where the stationarity assumption may be questionable. We leave these topics for future investigation.
	
	\vspace{-3mm}
	\section*{Acknowledgments}
	\vspace{-3mm}
	{\small This work was supported by the National Research Foundation of Korea(NRF) funded by the Korea government (2021R1A2C1091357; RS-2026-25471513; RS-2021-NR060140).}
	%
	\vspace{-3mm}
	\section*{Data Availability Statement}
	\vspace{-3mm}
	The data in this paper are available at \url{https://github.com/qsoon/gCChA}.

	\begin{appendices}
		\numberwithin{equation}{section}
		\numberwithin{theorem}{section}
		\numberwithin{assumption}{section}
		\numberwithin{lemma}{section}
		\numberwithin{figure}{section}
		\numberwithin{table}{section}
		\numberwithin{algorithm}{section}
		\def\thesection{\Alph{section}}
		
	\section{Error Minimization Approaches} \label{sec:error_minimize}
	We can approach the gCChA procedure from an error minimization perspective. To this end, we seek to approximate $Y_i$ using a sum of graph-filtered outputs of $Z_1, \ldots, Z_r$, 
	\begin{equation} \label{eq:gcca_reconstruct} 
		\mu_i + \mr{G}_{i1}Z_1 + \cdots + \mr{G}_{ir}Z_r =: \mu_i+\mc{T}_i^{\,\mr{G}}(Z), \quad 1 \le i \le q,
	\end{equation}
	where $\mu_i$ is a vector of length $n$ and $\mr{G}_{ij}=V \operatorname{diag}(\{g_{ij}(\lambda_\ell)\}_{\ell=1}^{n}) V^H$ is a graph filter. Substituting the expression of $Z_i$ in (1) into (\ref{eq:gcca_reconstruct}), we obtain
	\begin{equation*} \label{eq:yhat_final}
		\mu_i+\mr{A}_{i1}X_1+\cdots+\mr{A}_{ip}X_p =: \mu_i + \mc{T}_i^{\,\mr{A}}(X) , \qquad 1 \le i \le q,
	\end{equation*}
	where $\mr{A}_{ij}=\mr{G}_{i1} \mr{H}_{1j} + \cdots + \mr{G}_{ir} \mr{H}_{rj}$ is a graph filter with frequency responses, $a_{ij}(\lambda_\ell)=g_{i1}(\lambda_\ell) h_{1j}(\lambda_\ell) + \cdots + g_{ir}(\lambda_\ell) h_{rj}(\lambda_\ell)$, i.e., $\mr{A}_{ij}=V \operatorname{diag}(\{a_{ij}(\lambda_\ell)\}_{\ell=1}^{n}) V^H$. We aim to minimize the mean squared error $\sum_{i=1}^{q} E[(Y_i-(\mu_i + \mc{T}_i^{\,\mr{A}}(X)))^H(Y_i-(\mu_i + \mc{T}_i^{\,\mr{A}}(X)))]$, where the solution is given by the following theorem.
	
	\begin{theorem} \label{thm:gcca_one}
		Let $XY=(X_1 \mid \cdots \mid X_p \mid Y_1 \mid \cdots \mid Y_q)$ be a $(p+q)$-dimensional stationary graph signal with respect to the graph shift operator $S=V\Lambda V^H$ on $\mc{G}$. The means of $X_i$ and $Y_j$ are denoted by $\mu_i^X$ and $\mu_j^Y$, respectively. Let $p_{ij}^X$ and $p_{ij}^Y$ denote the graph cross-spectral densities of ($X_i, X_j$) and ($Y_i, Y_j$), respectively, and let $p_{ij}^{XY}$ and $p_{ij}^{YX}$ denote the graph cross-spectral densities of ($X_i, Y_j$) and ($Y_i, X_j$), respectively. For each graph frequency $\lambda_\ell$ ($1 \le \ell \le n$), define the spectral matrices,
		\[P_X(\lambda_\ell) \in \mathbb{C}^{p \times p},\;
		P_Y(\lambda_\ell) \in \mathbb{C}^{q \times q}, \; 
		P_{XY}(\lambda_\ell) \in \mathbb{C}^{p \times q}, \;
		P_{YX}(\lambda_\ell) \in \mathbb{C}^{q \times p},\]
		whose $(i,j)$th elements are $p_{ij}^X(\lambda_\ell)$, $p_{ij}^Y(\lambda_\ell)$, $p_{ij}^{XY}(\lambda_\ell)$, and $p_{ij}^{YX}(\lambda_\ell)$, respectively. Suppose that $P_X(\lambda_\ell)$ and $P_Y(\lambda_\ell)$ are nonsingular for all $1\le \ell \le n$. 
		{\color{black}
			Then, $\{\mu_i\}_{i=1}^q$ and $\{\mr{A}_{ij}\}_{1\le i \le q, 1\le j \le p}$ that minimize the mean squared error $\sum_{i=1}^{q} E[(Y_i-(\mu_i + \mc{T}_i^{\,\mr{A}}(X)))^H(Y_i-(\mu_i + \mc{T}_i^{\,\mr{A}}(X)))]$, are given by 
			\begin{equation*}
				\mu_i = \mu_i^Y -\sum_{j=1}^r \mr{G}_{ij}\left(\sum_{k=1}^p \mr{H}_{jk}\mu_k^X\right)~~\mbox{ and }~~
				A(\lambda_\ell) = G(\lambda_\ell) H(\lambda_\ell),
			\end{equation*}
			where  $H(\lambda_\ell) = \left(u_1(\lambda_\ell)|\cdots|u_r(\lambda_\ell)\right)^H P_{YX}(\lambda_\ell) P_X(\lambda_\ell)^{-1}$ and $G(\lambda_\ell) = \left(u_1(\lambda_\ell)|\cdots|u_r(\lambda_\ell)\right)$. Here, the matrices $A(\lambda_\ell)$, $H(\lambda_\ell)$, and $~G(\lambda_\ell)$ are of size $q \times p$, $r \times p$, and $q \times r$, respectively, and their $(i,j)$th elements are $a_{ij}(\lambda_\ell)$, $h_{ij}(\lambda_\ell)$, and  $g_{ij}(\lambda_\ell)$.
		} 
		Also $u_i(\lambda_\ell)$ denotes the $i$th orthonormal eigenvector of $P_{YX}(\lambda_\ell) P_X(\lambda_\ell)^{-1} P_{XY}(\lambda_\ell)$ associated with the $i$th eigenvalue $\tau_i(\lambda_\ell)$ (with $\tau_1(\lambda_\ell) \ge \cdots \ge \tau_q(\lambda_\ell) \ge 0$).  The minimum mean squared error is obtained as $\sum_{\ell=1}^{n} \operatorname{tr}[P_Y(\lambda_\ell) - P_{YX}(\lambda_\ell)P_X(\lambda_\ell)^{-1}P_{XY}(\lambda_\ell)] + \sum_{\ell=1}^{n} \sum_{i>r} \tau_i(\lambda_\ell)$.
	\end{theorem}
	
	\begin{proof}
		A proof is provided in Appendix~B.
	\end{proof}
	
	If $X=Y$, we obtain PCA results discussed in \citet{kim2024principal}.
	
	The above approach minimizes $\sum_{i=1}^{q} E[\epsilon_i^H\epsilon_i] = \sum_{\ell=1}^{n}\operatorname{tr}[P_\epsilon(\lambda_\ell)]$, where $\epsilon = (\epsilon_1 \mid \cdots \mid \epsilon_q)$ is a $q$-dimensional error signal with $\epsilon_i = Y_i-(\mu_i + \mc{T}_i^{\,\mr{A}}(X))$, and $P_\epsilon(\lambda_\ell)$ denotes the spectral matrix of $\epsilon$ at graph frequency $\lambda_\ell$.
	This criterion assigns equal weight to errors across all nodes, which may be inappropriate when signal variances differ across nodes. As an alternative, we may consider minimizing
	\begin{equation} \label{eq:alternative_loss}
		\sum_{\ell=1}^{n}\operatorname{tr}[P_Y(\lambda_\ell)^{-1/2}P_\epsilon(\lambda_\ell)P_Y(\lambda_\ell)^{-1/2}],    
	\end{equation}
	which leads to the solution stated in \Cref{cor:gcca_one}.
	\begin{corollary} \label{cor:gcca_one}
		Under the conditions of \Cref{thm:gcca_one}, the solution minimizing (\ref{eq:alternative_loss}) is determined by the eigenvectors of $P_Y(\lambda_\ell)^{-1/2}P_{YX}(\lambda_\ell) P_X(\lambda_\ell)^{-1} P_{XY}(\lambda_\ell)P_Y(\lambda_\ell)^{-1/2}$. 
	\end{corollary} 
	
	As a closely related problem, we consider minimizing  $\sum_{i=1}^r E[(W_i - \mu_i - Z_i)^H(W_i - \mu_i - Z_i)]$ subject to the constraints $F(\lambda_\ell)P_Y(\lambda_\ell)F(\lambda_\ell)^H=I_r$ and $H(\lambda_\ell)P_X(\lambda_\ell)H(\lambda_\ell)^H=I_r$, where $H(\lambda_\ell)$ and $F(\lambda_\ell)$ are $r \times p $ and $r \times q$ matrices whose $(i,j)$th elements are $h_{ij}(\lambda_\ell)$ and $f_{ij}(\lambda_\ell)$, respectively. The corresponding solution is given in \Cref{thm:gcca_sym_minimize}.

	\begin{theorem} \label{thm:gcca_sym_minimize}
		Under the conditions of \Cref{thm:gcca_one}, the solutions that minimize $\sum_{i=1}^r E[(W_i - \mu_i - Z_i)^H(W_i - \mu_i - Z_i)]$ subject to $F(\lambda_\ell)P_Y(\lambda_\ell)F(\lambda_\ell)^H=I_r$ and $H(\lambda_\ell)P_X(\lambda_\ell)H(\lambda_\ell)^H=I_r$ are given by
		\begin{eqnarray*}
			\mu_i &=& \sum_{j=1}^q \mr{F}_{ij}\mu_j^Y - \sum_{j=1}^p \mr{H}_{ij}\mu_j^X, \quad 1 \le i \le r,\\
			H(\lambda_\ell) &=& \left(
			d_1(\lambda_\ell)|  
			\cdots|
			d_r(\lambda_\ell)\right)^H P_X(\lambda_\ell)^{-1/2}, \\
			F(\lambda_\ell) &=& \left(
			e_1(\lambda_\ell) |  
			\cdots |
			e_r(\lambda_\ell)\right)^H P_Y(\lambda_\ell)^{-1/2},
		\end{eqnarray*}
		where the matrices $H(\lambda_\ell)$ and $F(\lambda_\ell)$ are of sizes $r \times p$ and $r \times q$, respectively, and their $(i,j)$th elements are  $h_{ij}(\lambda_\ell)$ and $f_{ij}(\lambda_\ell)$. Here,  $d_i(\lambda_\ell)$ is the $i$th orthonormal  eigenvector of $P_X(\lambda_\ell)^{-1/2} P_{XY}(\lambda_\ell)P_{Y}(\lambda_\ell)^{-1}P_{YX}(\lambda_\ell)P_{X}(\lambda_\ell)^{-1/2}$ and $e_i(\lambda_\ell)$ is the $i$th orthonormal  eigenvector of  $P_Y(\lambda_\ell)^{-1/2} P_{YX}(\lambda_\ell)P_{X}(\lambda_\ell)^{-1}P_{XY}(\lambda_\ell)P_{Y}(\lambda_\ell)^{-1/2}$. These two matrices share the same eigenvalues, and $d_i(\lambda_\ell)$ and $e_i(\lambda_\ell)$ correspond to their common $i$th largest eigenvalue.
	\end{theorem}

	\begin{proof}
		A proof is provided in Appendix~B.
	\end{proof}
	
	The resulting $Z_i$ and $W_i$ are equivalent to the solutions in Theorem 3.1.

	\section{Proofs} \label{sec:FA_PCA}
	
	\subsection{Proof of Theorem 3.1} \label{pf:gcca_sym}
	\begin{proof}
		We have $\Sigma_{ii}^Z = \mathbb{E}[(\sum_{j=1}^{p} \mr{H}_{ij}X_j)(\sum_{k=1}^{p} \mr{H}_{ik}X_k)^H] = \sum_{1\le j,k \le p} \mr{H}_{ij} \Sigma_{jk}^X \mr{H}_{ik}^H$. Then, we can easily derive $p_{ii}^Z(\lambda_\ell) = h_i(\lambda_\ell)^H P_X(\lambda_\ell) h_i(\lambda_\ell)$. Similarly, we can get $p_{ii}^W(\lambda_\ell) = g_i(\lambda_\ell)^H P_Y(\lambda_\ell) g_i(\lambda_\ell)$ and $p_{ii}^{ZW}(\lambda_\ell) = h_i(\lambda_\ell)^H P_{XY}(\lambda_\ell) g_i(\lambda_\ell)$. Therefore, the coherence of $Z_i$ and $W_i$ at $\lambda_\ell$, defined by $c_{ii}^{ZW}(\lambda_\ell) = \lvert p_{ii}^{ZW}(\lambda_\ell) \rvert^2/(p_{ii}^{Z}(\lambda_\ell)p_{ii}^{W}(\lambda_\ell))$, satisfies the following inequalities for $h_i'(\lambda_\ell) = P_X(\lambda_\ell)^{1/2}h_i(\lambda_\ell)$ and $f_i'(\lambda_\ell) = P_Y(\lambda_\ell)^{1/2}f_i(\lambda_\ell)$. 
		\begin{equation*}
			\begin{aligned}
				c_{ii}^{ZW}(\lambda_\ell)&=\frac{\lvert h_i'(\lambda_\ell)^H P_X(\lambda_\ell)^{-1/2} P_{XY}(\lambda_\ell) P_Y(\lambda_\ell)^{-1/2} f_i'(\lambda_\ell) \rvert^2}{(h_i'(\lambda_\ell)^H h_i'(\lambda_\ell))(f_i'(\lambda_\ell)^H f_i'(\lambda_\ell))}  \\
				& \le \frac{f_i'(\lambda_\ell)^H (P_Y(\lambda_\ell)^{-1/2} P_{YX}(\lambda_\ell) P_X(\lambda_\ell)^{-1} P_{XY}(\lambda_\ell) P_Y(\lambda_\ell)^{-1/2}) f_i'(\lambda_\ell)}{f_i'(\lambda_\ell)^H f_i'(\lambda_\ell)} \\
				& \le \gamma_i(\lambda_\ell).
			\end{aligned}
		\end{equation*}
		The first inequality holds by Cauchy--Schwarz inequality. The second inequality follows from the maximization property of quadratic forms. The equation in the second inequality is obtained when $f_i'(\lambda_\ell)$ is proportional to  $\eta_i(\lambda_\ell)$. The equation in the first inequality is obtained when $h_i'(\lambda_\ell)$ is proportional to  $P_X(\lambda_\ell)^{-1/2} P_{XY}(\lambda_\ell) P_Y(\lambda_\ell)^{-1/2} \eta_i(\lambda_\ell)$. Therefore, the solutions $h_i(\lambda_\ell)$ and  $f_i(\lambda_\ell)$ are proportional to $P_X(\lambda_\ell)^{-1} P_{XY}(\lambda_\ell)P_{Y}(\lambda_\ell)^{-1/2}\eta_i(\lambda_\ell)$ and $P_Y(\lambda_\ell)^{-1/2} \eta_i(\lambda_\ell)$, respectively. From the above inequality, the maximum coherence is given by $\gamma_i(\lambda_\ell)$. 
		
		We note that for the solutions $h_i(\lambda_\ell)$ and $f_i(\lambda_\ell)$, the associated quantity $h_i'(\lambda_\ell)$ is the normalized eigenvector of $$P_X(\lambda_\ell)^{-1/2} P_{XY}(\lambda_\ell) P_Y(\lambda_\ell)^{-1} P_{YX}(\lambda_\ell)P_X(\lambda_\ell)^{-1/2}$$ corresponding to the eigenvalue $\gamma_i(\lambda_\ell)$, and that   $f_i'(\lambda_\ell)$ is the  normalized $\eta_i(\lambda_\ell)$. It is clear that
		\begin{equation*}
			\begin{gathered}
				P_X(\lambda_\ell)^{-1} P_{XY}(\lambda_\ell)P_{Y}(\lambda_\ell)^{-1}P_{YX}(\lambda_\ell) h_i(\lambda_\ell) = \gamma_i(\lambda_\ell) h_i(\lambda_\ell),\\
				P_Y(\lambda_\ell)^{-1} P_{YX}(\lambda_\ell)P_{X}(\lambda_\ell)^{-1}P_{XY}(\lambda_\ell) f_i(\lambda_\ell) = \gamma_i(\lambda_\ell) f_i(\lambda_\ell).
			\end{gathered}
		\end{equation*}
		In addition, for these solutions, we have $p_{ii}^Z(\lambda_\ell) = h_i(\lambda_\ell)^HP_X(\lambda_\ell) h_i(\lambda_\ell)=1$ and $p_{ii}^W(\lambda_\ell) = f_i(\lambda_\ell)^H P_Y(\lambda_\ell)f_i(\lambda_\ell)=1$. Moreover, $p_{ij}^Z(\lambda_\ell) = h_i(\lambda_\ell)^H P_X(\lambda_\ell)h_j(\lambda_\ell) = h_i'(\lambda_\ell)^H h_j'(\lambda_\ell)$, which is equal to zero for $j<i$. Similarly, $p_{ij}^W(\lambda_\ell) = 0$ for $j<i$. It also holds that $p_{ij}^{ZW}(\lambda_\ell) = h_i(\lambda_\ell)^H P_{XY}(\lambda_\ell) f_j(\lambda_\ell) = h_i'(\lambda_\ell)^H P_X(\lambda_\ell)^{-1/2} P_{XY}(\lambda_\ell) P_Y(\lambda_\ell)^{-1/2} f_j'(\lambda_\ell)$ is proportional to $\eta_i(\lambda_\ell)^H P_Y(\lambda_\ell)^{-1/2} P_{YX}(\lambda_\ell) P_X(\lambda_\ell)^{-1} P_{XY}(\lambda_\ell) P_Y(\lambda_\ell)^{-1/2} \eta_j(\lambda_\ell) = 0$ for $j <i$. Similarly, $p_{ij}^{WZ}(\lambda_\ell) = 0$ for $j<i$. Therefore, the graph coherences between $Z_i$ and $Z_j$, $W_j$ for $j<i$ are zeros and those between $W_i$ and $Z_j$, $W_j$ for $j<i$ are zeros.
	\end{proof}
	

	\subsection{Proof of Proposition 3.1}
	\begin{proof}
		The coherence of $Z_i$ and $X_j$ at graph frequency $\lambda_\ell$ is defined as 
		\[c_{ij}^{ZX}(\lambda_\ell) = \frac{\lvert p_{ij}^{ZX}(\lambda_\ell) \rvert^2}{p_{ii}^Z(\lambda_\ell) p_{jj}^X(\lambda_\ell)}.\]
		For the canonical basis vector $e_j = (0,\ldots, 0, 1, 0,\ldots, 0)^\top$, this becomes
		\[c_{ij}^{ZX}(\lambda_\ell) = \frac{\lvert h_i(\lambda_\ell)^H P_X(\lambda_\ell) e_j \rvert^2}{e_j^H P_X(\lambda_\ell) e_j} = \frac{e_j^H P_X(\lambda_\ell) h_i(\lambda_\ell) h_i(\lambda_\ell)^H P_X(\lambda_\ell) e_j}{e_j^H P_X(\lambda_\ell) e_j}.\]
		Then, 
		\[\sum_{i=1}^r c_{ij}^{ZX}(\lambda_\ell) = \frac{e_j^H P_X(\lambda_\ell) H(\lambda_\ell)^H H(\lambda_\ell) P_X(\lambda_\ell) e_j}{e_j^H P_X(\lambda_\ell) e_j}.\]
		By the maximization property of quadratic forms, the maximum value of $\sum_{i=1}^r c_{ij}^{ZX}(\lambda_\ell)$ corresponds to the largest eigenvalue of $P_X(\lambda_\ell)^{1/2} H(\lambda_\ell)^H H(\lambda_\ell) P_X(\lambda_\ell)^{1/2}$. Since $H(\lambda_\ell)P_X(\lambda_\ell) H(\lambda_\ell)^H = I_r$, the matrix $P_X(\lambda_\ell)^{1/2} H(\lambda_\ell)^H H(\lambda_\ell) P_X(\lambda_\ell)^{1/2}$ is Hermitian and idempotent. Therefore, its eigenvalues are either  0 or 1, and hence, 
		\[\sum_{i=1}^r c_{ij}^{ZX}(\lambda_\ell) \le 1.\]
		Similarly, 
		\[\sum_{i=1}^r c_{ij}^{WY}(\lambda_\ell) \le 1.\]
		Consequently, 
		\[
		\begin{gathered}
			\frac{1}{p}\sum_{i=1}^r \sum_{j=1}^p c_{ij}^{ZX}(\lambda_\ell) \le 1, \quad
			\frac{1}{q}\sum_{i=1}^r \sum_{j=1}^q c_{ij}^{WY}(\lambda_\ell) \le 1.
		\end{gathered}\]
		The non-negativity is obvious.
		
		In the special case where $r = \min(p, q)$, consider first the case $p > q$ ($r = q$).
		From $F(\lambda_\ell) P_Y(\lambda_\ell) F(\lambda_\ell)^H= I_q$, it follows that $P_Y(\lambda_\ell)^{1/2} F(\lambda_\ell)^H F(\lambda_\ell) P_Y(\lambda_\ell)^{1/2} = I_q$. Thus, $\sum_{i=1}^r c_{ij}^{WY}(\lambda_\ell) = 1$. Similarly, if $p<q$ ($r=p$), then $\sum_{i=1}^r c_{ij}^{ZX}(\lambda_\ell) = 1$.
	\end{proof}

	\subsection{Proof of \Cref{thm:gcca_one}} \label{pf:gcca_one}
	\begin{proof}
		Denote the error to be minimized by $J$. Then, using (\ref{eq:gcca_reconstruct}), the error becomes
		\begin{equation*}
			\begin{aligned}
				J &= \sum_{i=1}^{q} E[(Y_i-(\mu_i+\mc{T}_i^{\,\mr{G}}(Z)))^H(Y_i-(\mu_i+\mc{T}_i^{\,\mr{G}}(Z)))] \\
				&= \sum_{i=1}^{q} \operatorname{tr}[E[(Y_i-(\mu_i+\mc{T}_i^{\,\mr{G}}(Z)))(Y_i-(\mu_i+\mc{T}_i^{\,\mr{G}}(Z)))^H]] \\
				&= \sum_{i=1}^{q} \operatorname{tr}[E[\{(Y_i-\mu_i^Y -\sum_{j=1}^r \mr{G}_{ij}(Z_j-\mu_j^Z)) + (\mu_i^Y-\mu_i -\sum_{j=1}^r \mr{G}_{ij}\mu_j^Z)\}  \\
				&\quad \quad \quad \quad \quad \quad \{(Y_i-\mu_i^Y -\sum_{j=1}^r \mr{G}_{ij}(Z_j-\mu_j^Z)) + (\mu_i^Y-\mu_i -\sum_{j=1}^r \mr{G}_{ij}\mu_j^Z)\}^H]],
			\end{aligned}
		\end{equation*}
		where $Z_j$ is defined by (1), and $\mu_j^Z = E[Z_j] = \sum_{k=1}^p \mr{H}_{jk} \mu_k^X$. Note that $\sum_{j=1}^r \mr{G}_{ij}(Z_j-\mu_j^Z) = \sum_{j=1}^r \mr{G}_{ij} (\sum_{k=1}^p \mr{H}_{jk}(X_k-\mu_k^X)) = \sum_{j=1}^p \mr{A}_{ij}(X_j-\mu_j^X)$. Consequently,  
		\begin{align} \label{eq:twoterms}
			J & = \sum_{i=1}^{q} \operatorname{tr}[E[\{(Y_i-\mu_i^Y -\sum_{j=1}^p \mr{A}_{ij}(X_j - \mu_j^X)) + (\mu_i^Y-\mu_i -\sum_{j=1}^r \mr{G}_{ij}\mu_j^Z)\} \nonumber  \\
			&\quad \quad \quad \quad \quad \quad \{(Y_i-\mu_i^Y -\sum_{j=1}^p \mr{A}_{ij}(X_j - \mu_j^X)) + (\mu_i^Y-\mu_i -\sum_{j=1}^r \mr{G}_{ij}\mu_j^Z)\}^H]] \nonumber\\
			&= \sum_{i=1}^{q} \operatorname{tr}[\Sigma_{ii}^Y - \sum_{j=1}^p \mr{A}_{ij}\Sigma_{ji}^{XY} - \sum_{j=1}^p \Sigma_{ij}^{YX}\mr{A}_{ij}^H + \sum_{1\le j,k \le p}\mr{A}_{ij}\Sigma_{jk}^X\mr{A}_{ik}^H]  \nonumber\\
			& \quad \quad \quad \quad \quad \quad + \; \sum_{i=1}^{q} (\mu_i^Y-\mu_i -\sum_{j=1}^r \mr{G}_{ij}\mu_j^Z)^H (\mu_i^Y-\mu_i -\sum_{j=1}^r \mr{G}_{ij}\mu_j^Z) \nonumber\\
			&= \sum_{\ell=1}^n \underbrace{\left[ \sum_{i=1}^{q} \left\{p_{ii}^Y(\lambda_\ell) - \sum_{j=1}^p a_{ij}(\lambda_\ell)p_{ji}^{XY}(\lambda_\ell) - \sum_{j=1}^p p_{ij}^{YX}(\lambda_\ell)a_{ij}^*(\lambda_\ell) + \sum_{1 \le j,k \le p}a_{ij}(\lambda_\ell)p_{jk}^X(\lambda_\ell)a_{ik}^*(\lambda_\ell)  \right\}\right]}_{J(\lambda_\ell)} \nonumber\\
			& \quad \quad \quad \quad \quad \quad + \;\sum_{i=1}^{q} (\mu_i^Y-\mu_i -\sum_{j=1}^r \mr{G}_{ij}\mu_j^Z)^H (\mu_i^Y-\mu_i -\sum_{j=1}^r \mr{G}_{ij}\mu_j^Z),
		\end{align}
		where $\mr{A}_{ij}=V \operatorname{diag}(\{a_{ij}(\lambda_\ell)\}_{\ell=1}^{n}) V^H$, and $\Sigma_{ij}^X$, $\Sigma_{ij}^Y$, $\Sigma_{ij}^{XY}$, and $\Sigma_{ij}^{YX}$ are the cross-covariance matrices between $X_i$ and $X_j$, $Y_i$ and $Y_j$, $X_i$ and $Y_j$, and $Y_i$ and $X_j$, respectively. If we compute the solutions  $\mr{H}_{ij}$ and $\mr{G}_{ij}$ that minimize $J$, we obtain the solution
		\begin{equation*}
			\mu_i =  \mu_i^Y -\sum_{j=1}^r \mr{G}_{ij}\mu_j^Z = \mu_i^Y -\sum_{j=1}^r \mr{G}_{ij}\left(\sum_{k=1}^p \mr{H}_{jk}\mu_k^X\right),
		\end{equation*} which makes the latter term on the right-hand side of (\ref{eq:twoterms}) equal to zero. 
		
		Now, let us find the solutions that minimize the first term on the right-hand side of (\ref{eq:twoterms}). To achieve this, we need to minimize $J(\lambda)$ for each $\lambda_\ell$ ($1 \le \ell \le n$). Define the matrices $A(\lambda_\ell)$, $H(\lambda_\ell)$, and $G(\lambda_\ell)$ as  $q \times p$, $r \times p$, and $q \times r$ matrices, respectively, where the $(i,j)$th elements are $a_{ij}(\lambda_\ell)$, $h_{ij}(\lambda_\ell)$, and  $g_{ij}(\lambda_\ell)$, representing the frequency responses of the graph filters $\mr{A}_{ij}$, $\mr{H}_{ij}$, and $\mr{G}_{ij}$, respectively. Then, we have $A(\lambda_\ell) = G(\lambda_\ell)H(\lambda_\ell)$, implying that $rank(A(\lambda_\ell)) \le r$. Moreover, Corollary 3.1 in \citet{kim2024principal} ensures that $P_X^{1/2}(\lambda_\ell)$ exists. With this in mind, $J(\lambda_\ell)$ becomes 
		\begin{equation*}
			\begin{aligned}
				J(\lambda_\ell)&=\operatorname{tr}[P_Y(\lambda_\ell) - A(\lambda_\ell) P_{XY}(\lambda_\ell) - P_{YX}(\lambda_\ell)A(\lambda_\ell)^H +A(\lambda_\ell)P_X(\lambda_\ell)A(\lambda_\ell)^H]  \\
				& = \operatorname{tr}[P_Y(\lambda_\ell) - P_{YX}(\lambda_\ell)P_X(\lambda_\ell)^{-1}P_{XY}(\lambda_\ell)] \\
				& \quad \quad  +  \operatorname{tr}[(P_{YX}(\lambda_\ell)P_X(\lambda_\ell)^{-1/2} - A(\lambda_\ell)P_X(\lambda_\ell)^{1/2})(P_{YX}(\lambda_\ell)P_X(\lambda_\ell)^{-1/2} - A(\lambda_\ell)P_X(\lambda_\ell)^{1/2})^H].
			\end{aligned}
		\end{equation*} 
		The $i$th eigenvectors of $P_{YX}(\lambda_\ell)P_X(\lambda_\ell)^{-1}P_{XY}(\lambda_\ell)$ and $P_X(\lambda_\ell)^{-1/2}P_{XY}(\lambda_\ell)P_{YX}(\lambda_\ell)P_X(\lambda_\ell)^{-1/2}$ are denoted by $u_i(\lambda_\ell)$ and $v_i(\lambda_\ell)$, respectively. Nonzero eigenvalues of the two matrices are equal by Theorem 3.7.2 of \citet{Brillinger2001time} and we denote the $i$th eigenvalue of $P_{YX}(\lambda_\ell)P_X(\lambda_\ell)^{-1}P_{XY}(\lambda_\ell)$ by $\tau_i(\lambda_\ell)$. Thus, by Corollary 3.7.4 of \citet{Brillinger2001time}, $J(\lambda_\ell)$ is minimized when 
		\begin{equation*}
			A(\lambda_\ell)P_X(\lambda_\ell)^{1/2} = \sum_{i=1}^r \tau_i(\lambda_\ell)^{1/2} u_i(\lambda_\ell)v_i(\lambda_\ell)^H.
		\end{equation*}
		Therefore, the indicated $A(\lambda_\ell) = G(\lambda_\ell) H(\lambda_\ell)$ minimizes $J(\lambda_\ell)$. 
		
		For the minimizing solution discussed above, we have $P_{YX}(\lambda_\ell)P_X(\lambda_\ell)^{-1/2} - A(\lambda_\ell)P_X(\lambda_\ell)^{1/2} = \sum_{i>r} \tau_i(\lambda_\ell)^{1/2} u_i(\lambda_\ell)v_i(\lambda_\ell)^H$. Therefore, $J(\lambda_\ell)$ becomes $\operatorname{tr}[P_Y(\lambda_\ell) - P_{YX}(\lambda_\ell)P_X(\lambda_\ell)^{-1}P_{XY}(\lambda_\ell)] + \sum_{i>r} \tau_i(\lambda_\ell)$. Consequently, the minimum mean squared error is obtained as $\sum_{\ell=1}^{n} \operatorname{tr}[P_Y(\lambda_\ell) - P_{YX}(\lambda_\ell)P_X(\lambda_\ell)^{-1}P_{XY}(\lambda_\ell)] + \sum_{\ell=1}^{n} \sum_{i>r} \tau_i(\lambda_\ell)$. This completes the proof.
	\end{proof}

	\subsection{Proof of \Cref{thm:gcca_sym_minimize}} \label{pf:gcca_sym_minimize}
	\begin{proof}
		Denote the error to be minimized by $J_0$. Then, the error becomes
		\begin{equation*}
			\begin{aligned}
				J_0 &= \sum_{i=1}^{r} E[(W_i-\mu_i -Z_i)^H(W_i-\mu_i -Z_i)] = \sum_{i=1}^{r} \operatorname{tr}[E[(W_i-\mu_i -Z_i)(W_i-\mu_i -Z_i)^H]] \\
				&= \sum_{i=1}^{r} \operatorname{tr}[E[\{(W_i-\mu_i^W -(Z_i-\mu_i^Z)) + (\mu_i^W-\mu_i -\mu_i^Z)\}  \\
				&\quad \quad \quad \quad \quad \quad \{(W_i-\mu_i^W -(Z_i-\mu_i^Z)) + (\mu_i^W-\mu_i -\mu_i^Z)\}^H]],
			\end{aligned}
		\end{equation*}
		where $\mu_i^Z = E[Z_i] = \sum_{j=1}^p \mr{H}_{ij} \mu_j^X$ and $\mu_i^W = E[W_i] = \sum_{j=1}^p \mr{F}_{ij} \mu_j^Y$. Similar to the proof of \Cref{thm:gcca_one},  
		\begin{align} \label{eq:twoterms2}
			J_0 & = \sum_{i=1}^{r} \operatorname{tr}\left[\sum_{1\le j,k \le q} \mr{F}_{ij}\Sigma_{jk}^Y \mr{F}_{ik}^H- \sum_{1\le j\le p, \,1\le k \le q} \mr{H}_{ij}\Sigma_{jk}^{XY}\mr{F}_{ik}^H - \sum_{1\le j\le p, \,1\le k \le q} \mr{F}_{ik}\Sigma_{kj}^{YX}\mr{H}_{ij}^H + \sum_{1\le j,k \le p} \mr{H}_{ij}\Sigma_{jk}^X \mr{H}_{ik}^H \right]  \nonumber\\
			& \quad \quad \quad \quad \quad \quad + \; \sum_{i=1}^{r} \left(\sum_{j=1}^q \mr{F}_{ij}\mu_j^Y-\mu_i -\sum_{j=1}^p \mr{H}_{ij}\mu_j^X \right)^H \left(\sum_{j=1}^q \mr{F}_{ij}\mu_j^Y-\mu_i -\sum_{j=1}^p \mr{H}_{ij}\mu_j^X\right) \nonumber\\
			&= \sum_{\ell=1}^n [ \sum_{i=1}^{r} \{ \sum_{1\le j,k \le q} f_{ij}(\lambda_\ell)p_{jk}^Y(\lambda_\ell)f_{ik}^*(\lambda_\ell) - \sum_{1\le j\le p, \,1\le k \le q}  h_{ij}(\lambda_\ell)p_{jk}^{XY}(\lambda_\ell)f_{ik}^*(\lambda_\ell) \nonumber \\
			& \quad \quad \quad \quad \quad \quad - \sum_{1\le j\le p, \,1\le k \le q} f_{ik}(\lambda_\ell) p_{ij}^{YX}(\lambda_\ell)h_{ij}^*(\lambda_\ell) + \sum_{1 \le j,k \le p} h_{ij}(\lambda_\ell) p_{jk}^X(\lambda_\ell)h_{ik}^*(\lambda_\ell) \}] \nonumber\\
			& \quad \quad \quad \quad \quad \quad + \;\sum_{i=1}^{r} \left(\sum_{j=1}^q \mr{F}_{ij}\mu_j^Y-\mu_i -\sum_{j=1}^p \mr{H}_{ij}\mu_j^X \right)^H \left(\sum_{j=1}^q \mr{F}_{ij}\mu_j^Y-\mu_i -\sum_{j=1}^p \mr{H}_{ij}\mu_j^X\right) \nonumber \\
			&=: \sum_{\ell=1}^n J_0(\lambda_\ell)+ \;\sum_{i=1}^{r} \left(\sum_{j=1}^q \mr{F}_{ij}\mu_j^Y-\mu_i -\sum_{j=1}^p \mr{H}_{ij}\mu_j^X \right)^H \left(\sum_{j=1}^q \mr{F}_{ij}\mu_j^Y-\mu_i -\sum_{j=1}^p \mr{H}_{ij}\mu_j^X\right).
		\end{align}
		If we compute the solutions  $\mr{H}_{ij}$ and $\mr{F}_{ij}$ that minimize $J_0$, we obtain the solution
		\begin{equation*}
			\mu_i =  \sum_{j=1}^q \mr{F}_{ij}\mu_i^Y -\sum_{j=1}^p \mr{H}_{ij}\mu_j^X,
		\end{equation*} which makes the latter term on the right-hand side of (\ref{eq:twoterms2}) equal to zero. 
		
		The expression $J_0(\lambda_\ell)$ can be reformulated as
		\begin{equation*}
			\begin{aligned}
				J_0(\lambda_\ell)&=\operatorname{tr}[F(\lambda_\ell)P_Y(\lambda_\ell)F(\lambda_\ell)^H - H(\lambda_\ell) P_{XY}(\lambda_\ell)F(\lambda_\ell)^H \\
				&\quad \quad \quad - F(\lambda_\ell)P_{YX}(\lambda_\ell)H(\lambda_\ell)^H +H(\lambda_\ell)P_X(\lambda_\ell)H(\lambda_\ell)^H].
			\end{aligned}
		\end{equation*} 
		The proof is then completed by following a similar procedure to that in Theorem 10.2.2 of \citet{Brillinger2001time}.
	\end{proof}
	
	\section{The gCChA Algorithm} 
	\begin{algorithm}
		\caption{gCChA}\label{alg:gcca}
		\begin{algorithmic}[1]
			\footnotesize
			\setlength{\itemsep}{2pt}
			\setlength{\parsep}{2pt}
			\State \textbf{Input:} $S$, $X_1, \ldots, X_p$, $Y_1, \ldots, Y_q$  
			\State \textbf{Output:} $\hat{Z}_1, \ldots, \hat{Z}_{p \wedge q}$, $\hat{W}_1, \ldots, \hat{W}_{p \wedge q}$, $\hat{\gamma}_1(\lambda_1), \ldots, \hat{\gamma}_1(\lambda_n), \ldots, \hat{\gamma}_{p \wedge q}(\lambda_1), \ldots, \hat{\gamma}_{p \wedge q}(\lambda_n)$ 
			\State $n \gets \text{the number of rows in } S$ 
			\State $r \gets p \wedge q$  \Comment{the number of canonical graph signals} 
			\State Perform eigendecomposition:  $S = V \Lambda V^H$ \Comment{$\lambda_1,\ldots, \lambda_n$: eigenvalues of $S$}
			\phase{Estimation of GCSD}
			\For{$\ell \gets 1$ \textbf{to} $n$}
			\State Estimate spectral matrices $P_X(\lambda_\ell)$, $P_Y(\lambda_\ell)$, $P_{XY}(\lambda_\ell)$, and $P_{YX}(\lambda_\ell)$
			\State and denote them by $\hat{P}_X(\lambda_\ell)$, $\hat{P}_Y(\lambda_\ell)$, $\hat{P}_{XY}(\lambda_\ell)$, and $\hat{P}_{YX}(\lambda_\ell)$
			\EndFor
			
			\phase{Evaluation of canonical graph signals}
			\State Estimate graph filters:
			
			\For{$\ell \gets 1$ \textbf{to} $n$}
			\State $\hat{h}_{i}(\lambda_\ell) \gets \text{the $i$th eigenvector of $\hat{P}_X(\lambda_\ell)^{-1} \hat{P}_{XY}(\lambda_\ell)\hat{P}_{Y}(\lambda_\ell)^{-1}\hat{P}_{YX}(\lambda_\ell)$}$
			\State $\hat{\gamma}_{i}(\lambda_\ell) \gets \text{the $i$th eigenvalue of $\hat{P}_X(\lambda_\ell)^{-1} \hat{P}_{XY}(\lambda_\ell)\hat{P}_{Y}(\lambda_\ell)^{-1}\hat{P}_{YX}(\lambda_\ell)$}$
			\State $\hat{f}_{i}(\lambda_\ell) \gets \text{the $i$th eigenvector of $\hat{P}_Y(\lambda_\ell)^{-1} \hat{P}_{YX}(\lambda_\ell)\hat{P}_{X}(\lambda_\ell)^{-1}\hat{P}_{XY}(\lambda_\ell)$}$
			\EndFor

			
			\State $\hat{\mr{H}}_{ij} \gets V\operatorname{diag}(\{\hat{h}_{ij}(\lambda_\ell)\}_{\ell=1}^{n})V^H, \quad 1 \le i \le r, ~1 \le j \le p$
			\State $\hat{\mr{F}}_{ij} \gets V\operatorname{diag}(\{\hat{f}_{ij}(\lambda_\ell)\}_{\ell=1}^{n})V^H, \quad 1 \le i \le r, ~1 \le j \le q$
			\State $\hat{Z}_i \gets \hat{\mr{H}}_{i1}X_1 + \cdots + \hat{\mr{H}}_{ip}X_p, \quad 1 \le i \le r$ \Comment{canonical graph signal of $X$}
			\State $\hat{W}_i \gets \hat{\mr{F}}_{i1}Y_1 + \cdots + \hat{\mr{F}}_{iq}Y_q, \quad 1 \le i \le r$ \Comment{canonical graph signal of $Y$}
		\end{algorithmic}
	\end{algorithm}
	
	\clearpage
	
	\section{Additional Tables} \label{sec:additional_figs_tbls}
	\begin{table}[!h]
		\captionsetup{justification=justified}
		\caption{Country codes and their corresponding countries in the trading network}
		\centering
		\renewcommand{\arraystretch}{1.4}
		\resizebox{0.75\textwidth}{!}{ 
			\begin{tabular}{cc @{\hskip 4em} cc @{\hskip 4em} cc}
				\hline
				Code & Country & Code & Country & Code & Country \\
				\hline
				AUS & Australia         & GBR & United Kingdom        & MEX & Mexico \\
				BRA & Brazil            & IDN & Indonesia             & RUS & Russian Federation \\
				CAN & Canada            & IND & India                 & SAU & Saudi Arabia \\
				CHN & China             & ITA & Italy                 & TUR & T\"{u}rkiye \\
				DEU & Germany           & JPN & Japan                 & USA & United States of America \\
				FRA & France            & KOR & Republic of Korea     & ZAF & South Africa \\
				\hline
			\end{tabular}
			\label{tbl:trading_countries}
		}
	\end{table}
	
	\begin{table}[!h]
		\captionsetup{justification=justified}
		\caption{List of economic and energy indicators used in the canonical graph analysis.}
		\centering
		\renewcommand{\arraystretch}{1.4}
		\resizebox{\textwidth}{!}{ 
			\begin{tabular}{cll}
				\hline
				\textbf{Signals} & \textbf{Indicators} & \textbf{Description} \\ \hline
				$X_1$ & Exports of goods and services & 
				Value of all goods and services provided to the rest of the world, including merchandise and various business services. \\
				
				$X_2$ & Foreign direct investment & 
				Net inflows of investment from abroad, including equity capital, reinvestment of earnings, and other capital. \\
				
				$X_3$ & GDP growth & 
				Annual percentage growth rate of GDP at market prices based on constant local currency. \\
				
				$X_4$ & Imports of goods and services & 
				Value of all goods and services received from abroad, including transport and business services. \\
				
				$X_5$ & Inflation & 
				Annual percentage change in the consumer price index, indicating inflation in consumer goods and services. \\ \hline
				
				$Y_1$ & Energy imports & 
				Net energy imports as gross imports minus gross exports (in tons of oil equivalent). \\
				
				$Y_2$ & Energy use & 
				Total primary energy consumption before transformation into other fuels. \\
				
				$Y_3$ & Fossil fuel consumption & 
				Consumption of coal, oil, petroleum, and natural gas products.\\
				
				$Y_4$ & GDP per unit of energy use & 
				GDP per unit of energy use, indicating energy efficiency. \\
				
				$Y_5$ & Renewable energy consumption & 
				Share of renewable energy, including hydro, solar, wind, and biofuels, in total final energy consumption. \\ \hline
			\end{tabular}
			\label{tbl:econ_energy_vars}
		}
	\end{table}
	
	\section{Classification Analysis of USPS Dataset}  \label{sec:classifiation_comparison}
\begin{figure}[!t]
		\centering
		\includegraphics[width=0.5\textwidth]{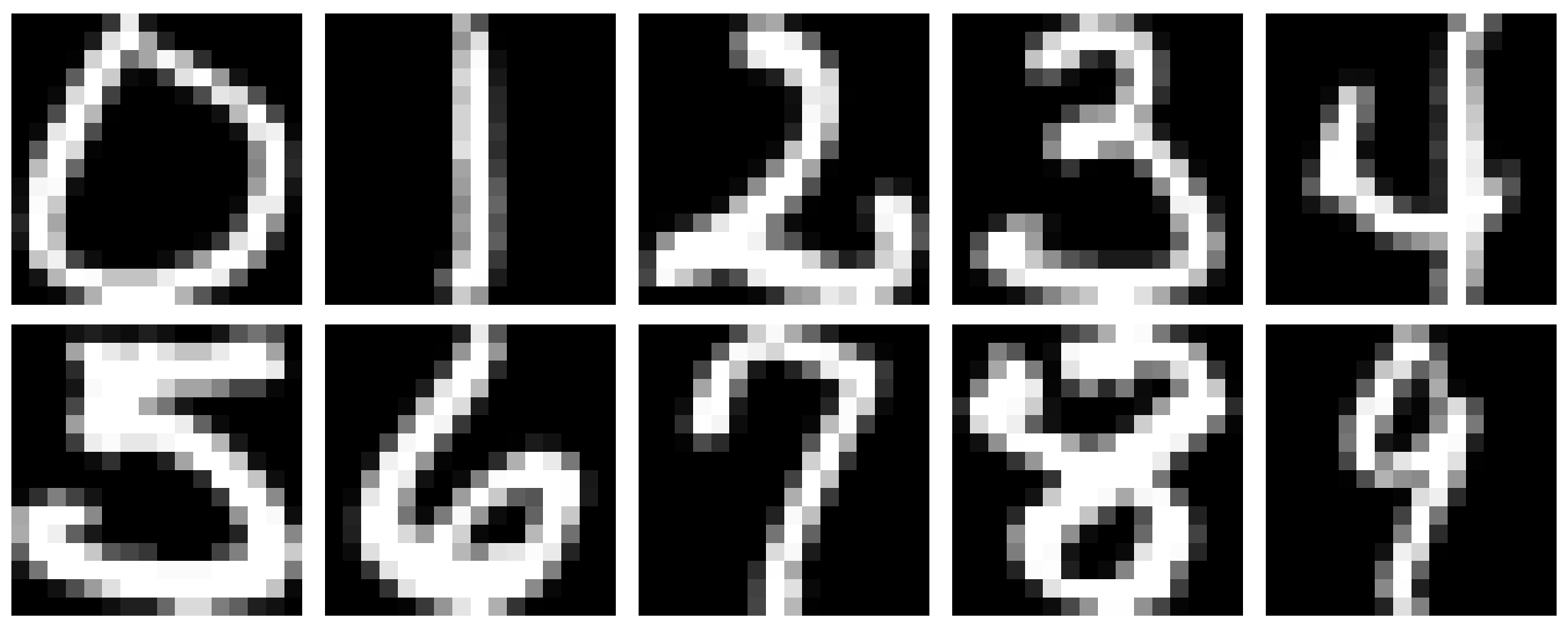}
		\vspace{-2mm}
		\caption{Illustration of the USPS dataset.}
		
		\label{fig:usps}
	\end{figure}

	We compare the proposed gCChA with the gCCA method of \citet{chen2018canonical} through an image classification task. \textcolor{black}{This experiment is included primarily to facilitate a direct comparison with gCCA.} For this purpose, we consider the USPS handwritten digit dataset \citep{Hull1994}, which consists of $16 \times 16$ grayscale images of digits from 0 to 9, as illustrated in \Cref{fig:usps}.
	
	We use a total of 400 images, with 40 images per digit class. Following the classification method in \citet{chen2018canonical}, we construct a graph where each image is a node. 
	Nodes sharing the same digit label are connected, with edge weights given by cosine similarities of the original 256-dimensional pixel vectors. We split the $256$ pixel features of each image into two parts: the top $K$ rows form a $(16 \times K)$-dimensional graph signal, and the remaining $(16-K)$ rows form a $(16 \times (16-K))$-dimensional graph signal. We consider five partitions with $K \in \{4,6,8,10,12\}$. For each image, these two parts define two graph signals.
	For each partition, we compute $r$ pairs of canonical graph signals ($r \in \{20,40\}$) and concatenate them into a $2r$-dimensional feature vector for each image. Classification is then performed on these feature vectors using a 10-nearest-neighbor classifier. We run the experiment 50 times. In each run, we randomly sample 400 images (40 per digit class), construct the graph, compute the canonical graph signals, and assess classification accuracy. The comparison is based on the average accuracy over 50 repetitions.
	
	Hyperparameter tuning for gCCA follows the procedure in \citet{chen2018canonical}, using logarithmically spaced values from $10^{-3}$ to $10^3$. We note an important conceptual difference between the two approaches: \citet{chen2018canonical} evaluated test accuracies by projecting test images onto the learned canonical vectors from the training set, whereas in gCChA, the graph filtering operations rely on the topology of the underlying graph. Because the graph constructed from test images generally differs from that constructed from training images, the filters learned from training cannot be directly applied to test images. In particular, when the training and test sample sizes differ, the corresponding matrices have mismatched dimensions, and filtering operations cannot be performed. As a result, the comparison between gCCA and gCChA is based on training accuracy. These results should not be interpreted as out-of-sample predictive performance. Rather, they are used as a within-sample measure of how well the canonical representations separate digit classes under the same graph construction.
	\Cref{tbl:classification_res} summarizes the classification performance. Across all proportions of the first part ($=K/16$), the proposed gCChA consistently achieves higher accuracy. These results suggest that gCChA can be effectively applied in image classification tasks.

	\begin{table}
		\captionsetup{justification=justified}
		\caption{USPS dataset classification results. Each cell shows the average classification accuracy across 50 repetitions, with standard deviations in parentheses.}
		\vspace{-2mm}
		\centering
		\resizebox{0.9\textwidth}{!}{ 
			\begin{tabular}{c|c|ccccc}
				\hline
				\multirow{2}{*}{$r$}  & \multirow{2}{*}{Methods} & \multicolumn{5}{c}{$K/16$}                                                      \\ \cline{3-7} 
				&                          & 0.25          & 0.375         & 0.5           & 0.625         & 0.75          \\ \hline
				\multirow{2}{*}{20}                  & gCCA                     & 0.979 (0.007) & 0.971 (0.007) & 0.967 (0.009) & 0.972 (0.008) & 0.980 (0.006) \\
				& gCChA                    & 0.992 (0.006) & 0.993 (0.005) & 0.993 (0.006) & 0.991 (0.007) & 0.993 (0.004) \\ \hline
				\multirow{2}{*}{40} & gCCA                     & 0.948 (0.009) & 0.923 (0.012) & 0.904 (0.017) & 0.913 (0.011) & 0.932 (0.011) \\ 
				& gCChA                    & 0.999 (0.001) & 0.999 (0.001) & 0.999 (0.000) & 0.999 (0.001) & 0.999 (0.001) \\ \hline
			\end{tabular}
			\label{tbl:classification_res}
		}
	\end{table}

	\end{appendices}
	
	
	\bibliographystyle{apalike}
	\bibliography{refs}
	
\end{document}